\documentclass[12pt]{iopart}
\usepackage{iopams}  

\usepackage{braket}

\usepackage{color}

\usepackage{graphicx}

\newcommand{\braketone}[1]{\langle#1\rangle}

\newcommand{\rb}{$^{87}$Rb }
\newcommand{\um}{$\mu$m }

\newcommand{\Teff}{T_{\mathrm{eff}}}
\newcommand{\gOneD}{g}

\definecolor{LinkColor}{rgb}{0,0,0.5}
\usepackage[%
	pdftitle={},%
	pdfauthor={},%
	pdfcreator={},
	pdfsubject={},
	pdfkeywords={}]{hyperref}
\hypersetup{colorlinks=true,%
	linkcolor=LinkColor,%
	citecolor=LinkColor,%
	filecolor=LinkColor,%
	menucolor=LinkColor,%
	urlcolor=LinkColor}
	
\begin{document}

\title{Prethermalization Revealed by the Relaxation Dynamics of Full Distribution Functions}

\author{D~Adu~Smith$^{1}$, M~Gring$^{1}$, T~Langen$^{1}$, M~Kuhnert$^{1}$, B~Rauer$^{1}$, R~Geiger$^{1}$, T~Kitagawa$^{2}$, I~Mazets$^{1,3}$, E~Demler$^{2}$ and J~Schmiedmayer$^{1}$}

\address{$^1$ Vienna Center for Quantum Science and Technology (VCQ), Atominstitut, TU~Wien, Vienna, Austria}
\address{$^2$ Harvard-MIT Center for Ultracold Atoms, Department of Physics, Harvard University, Cambridge, Massachusetts 02138, USA}
\address{$^3$ Ioffe Physico-Technical Institute of the Russian Acedemy of Sciences, 194021, St. Petersburg, Russia}

\ead{schmiedmayer@atomchip.org}

\begin{abstract} 
We detail the experimental observation of the non-equilibrium many-body phenomenon prethermalization. We study the dynamics of a rapidly and coherently split one-dimensional Bose gas. An analysis based on the use of full quantum mechanical probability distributions of matter wave interference contrast reveals that the system evolves towards a quasi-steady state. This state, which can be characterized by an effective temperature, is not the final thermal equilibrium state. We compare the evolution of the system to an integrable Tomonaga-Luttinger liquid model, and show that the system dephases to a prethermalized state rather than undergoing thermalization towards a final thermal equilibrium state. 
\end{abstract}

\tableofcontents

\section{Introduction}

A general understanding of the dynamics of non-equilibrium many-body quantum systems is an important unsolved problem impacting on many areas of physics~\cite{Polkovnikov2011,Lamacraft2011b}. In particular the question of why and how isolated quantum systems relax towards equilibrium states has only been studied for a very limited number of special systems~\cite{Deutsch1991,Srednicki1994,Rigol2008}. Moreover, in many cases, such as when dealing with integrable systems exhibiting many constants of motion, efficient relaxation can be completely absent~\cite{Kinoshita2006} or strongly inhibited. 

An intriguing phenomenon which has recently been experimentally demonstrated in this context is prethermalization~\cite{Berges2004,Gring2012}. The concept of prethermalization was first introduced in order to explain the remarkable success in applying thermodynamic models to heavy-ion collision experiments on time scales much shorter than the expected thermalization time of the scattered particles~\cite{Berges2004,BraunMunzinger2001,Rafelski2002,Rapp2004a}. In particular, the successful application of hydrodynamics~\cite{Heinz2002} relies on the assumption of being at least approximately close to thermal equilibrium at every point in space and the presence of an equation of state\,\cite{Ryblewski2011,Heinz2002a}.

Taking a low-energy quark-meson model as a model system, Berges et al.\,\cite{Berges2004} suggested in 2004 an explanation for this unexpected, apparent early thermalization. In their theoretical analysis, Berges et al. were able to show that after a very rapid initial evolution, their model system relaxes to a non-thermal quasi-steady state. This intermediate state is very robust with respect to changing initial conditions and already shows many of the bulk characteristics of the final thermal state. For example, the ratio of kinetic energy to pressure (i.e. the equation of state) or a properly defined kinetic temperature are already practically indistinguishable from their values in thermal equilibrium. In contrast, mode quantities such as the occupation numbers of different momentum modes are still far away from a true thermal equilibrium even though during the initial rapid evolution they also reach quasi-steady values. Their work showed that although the model system had not yet fully thermalized, it could in many aspects be described using a thermal model. Accordingly, the above described phenomenon was termed \textit{pre}-thermalization.

Since then, prethermalization has been discussed in the context of many other systems, ranging from models in relativistic field theories~\cite{Destri2006}, the dynamics of the early universe~\cite{Podolsky2006} to systems which can be realized using ultracold atoms, such as systems obeying the Hubbard model~\cite{Kollath2007, Eckstein2009,Moeckel2010}, spinor condensates~\cite{Barnett2010}, quantum Ising chains~\cite{Marino2012}, two-dimensional superfluids~\cite{Mathey2010} or one-dimensional (1d) quasi-condensates~\cite{Kitagawa2010,Kitagawa2011}.
In the present understanding, prethermalization is characterized by the rapid establishment of a quasi-stationary state on time scales much shorter than the expected thermalization time. This state is long-lived and already exhibits some thermal-like properties but can still be very different from the true thermal equilibrium state of the system. Full relaxation to thermal equilibrium, if present at all, is then expected to happen on a further much-longer time scale. While thermalization can be interpreted as a \textit{full} loss of the memory of the system about its initial state, prethermalization describes only a \textit{partial} loss of information about the initial state. 

In the following, we discuss the direct observation of the prethermalized state presented in~\cite{Gring2012}, giving, particularly, details of the experimental system and analysis methods used to reveal the prethermalization~\cite{Gring2012b}.

%
\section{Non-Equilibrium Dynamics of Interfering One-Dimensional Bose Gases}  \label{FDFs}
\subsection{Experimental System} 
The experimental study is performed using trapped 1d Bose gases. Such systems offer two unique advantages for non-equilibrium experiments: firstly, on the experimental side, realizing them with ultracold atoms facilitates a precise preparation and probing of the system. Secondly, on the theoretical side, 1d Bose gases offer a model system which contains complex many-body physics but can still be captured with reasonable theoretical effort, particularly due to the existence of effective models which allow to describe the essential physics in a relatively simple way~\cite{Giamarchi2004}. Furthermore, the homogeneous 1d Bose gas with repulsive contact interactions is an example of a fully integrable quantum system~\cite{Lieb1963,Yang1969}. The approximate realization of such a system in experiments thus allows the study of thermalization in the vicinity of multiple conserved quantities and hence the study of the interplay between integrability, many-body dynamics and thermalization.

\begin{figure}[hb]
\centering
\includegraphics[width=\textwidth]{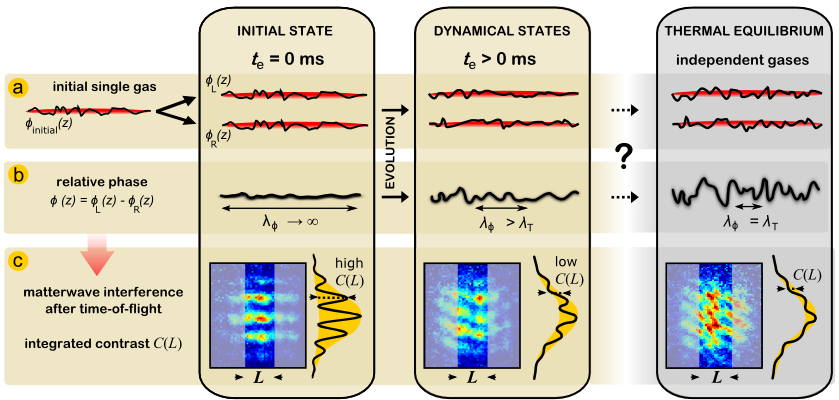}
\caption{Overview of the experiments performed in the context of this work~\cite{Gring2012}. (a,b) A phase fluctuating 1d quasi-condensate is coherently split creating two 1d gases with almost identical phase profiles $\phi_L(z)$ and $\phi_R(z)$ (represented by the black solid lines). The gases are then allowed to evolve in the double-well potential for some time $t_e$ which leads to strong fluctuations in the local phase difference $\phi(z)$ and a decrease of the phase correlation length $\lambda_{\phi}$. The question at which this experiment aims is whether and how this dynamical state reaches the thermal equilibrium state corresponding to two independently created quasi-condensates. In this thermal equilibrium state the phase difference between the 1d gases fluctuates strongly along its length and the correlation length $\lambda_T$ is determined by the temperature and density of each cloud. (c) The phase difference $\phi(z)$ between the two 1d gases is probed through time-of-flight matter wave interference of the two gases where the local relative phase is directly transformed to a local phase shift of the interference pattern. The contrast $C(L)$ of the axially integrated interference pattern can then be used as a direct measure of the strength of the relative phase fluctuations.}
\label{experiment schematics}
\end{figure}

Figure \ref{experiment schematics} summarizes the main idea of our experimental study of prethermalization using 1d Bose gases in the quasi-condensate regime~\cite{Petrov2000,Petrov2001,Kheruntsyan2003}. In this regime, density fluctuations are strongly suppressed and the gas is characterized by strong phase fluctuations. The properties of these phase fluctuations are determined by the temperature and the density of the system. In the experiment such a single trapped quasi-condensate is rapidly and coherently split, producing two uncoupled 1d gases with identical phase profiles. The aim of our study is to probe how these initially almost perfect correlations of the relative phase become obscured over time and if the thermal equilibrium state corresponding to two completely independent gases is finally reached~\cite{Hofferberth2008,Betz2011}. To this end, the two gases are allowed to evolve in the double-well potential for a varying evolution time $t_e$ before the relative phase correlations are  probed via time-of-flight matter wave interference (figure~\ref{experiment schematics}c). As differences in the relative phase lead to a locally displaced interference pattern, the contrast of the longitudinally integrated interference pattern is a direct probe for the dynamics of the system~\cite{Bistritzer2007,Polkovnikov2006,Imambekov2008,Schumm2005a}. Example interference patterns after various evolution times, demonstrating the loss of the initial phase coherence, are shown in figure~\ref{experiment schematics}. 


\subsection{Initial Non-Equilibrium State and Thermal Equilibrium}
A prerequisite for non-equilibrium experiments is the ability to precisely prepare and characterize both the initial non-equilibrium state, as well as the expected thermal equilibrium state of the system. One of the key advantages of coherently split 1d Bose gases is that both these states can be prepared and described with high precision. 

For a general system of two spatially separated 1d Bose gases, the excitations can be described by anti-symmetric and symmetric longitudinal modes which relate to the relative and common degrees of freedom of the two halves of the system. They are given by  
\begin{equation}
\phi(y) = \phi_L(y)-\phi_R(y) \quad\textrm{,}\quad \phi_\mathrm{com}(y)=\frac{\phi_R(y)+\phi_L(y)}{2}
\end{equation}
for the phase, and 
\begin{equation}
n(y)=\frac{n_R(y)-n_L(y)}{2} \quad\textrm{,}\quad n_\mathrm{com}(y)=n_R(y)+n_L(y)
\end{equation}
for the density. Here $\phi_{L,R}(y)$ describes the longitudinal phase profiles and $n_{L,R}(y)$ the density of the left and the right gas, respectively. 

\begin{figure}[tb]
\begin{center}
\includegraphics[width = 0.7\textwidth]{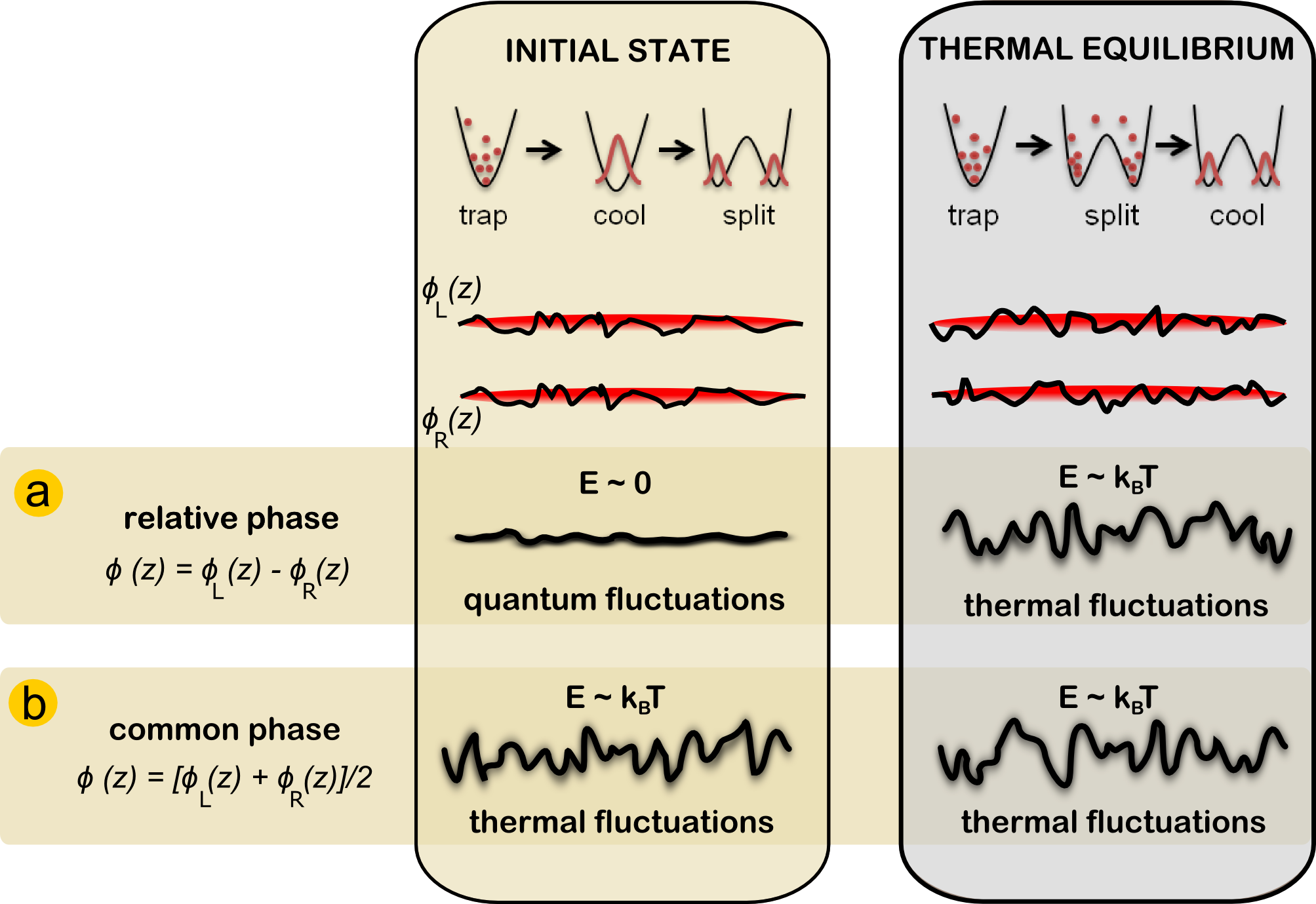}
\caption{Comparison of the initial non-equilibrium state prepared by the coherent splitting process and the thermal equilibrium state of two independent quasi-condensates. The system can be described by a combination of relative degrees of freedom (a) and common degrees of freedom (b). After the coherent splitting the common degrees of freedom contain all the thermal energy $E\sim k_B T$ of the system, whereas the relative degrees of freedom contain only the quantum noise of the splitting process. Here, $k_B$ is Boltzmann's constant and $T$ the temperature of the system. In thermal equilibrium all degrees of freedom contain the same amount of thermal energy and are thus characterized by the same temperature.}
\label{fig:states}
\end{center}
\end{figure}

After the coherent splitting, the individual phase profiles of the two halves are almost identical and hence the relative phase profile is almost flat (figure \ref{fig:states} left column). In terms of excitations, this means that all the thermal excitations are initially contained in the common degrees of freedom. As we will detail in section\,\ref{LL Description}, the relative degrees of freedom, on the other hand, are initially populated only by quantum noise created in the splitting process.
 
In thermal equilibrium we expect to find the system in a state where the energies in the relative and in the common degrees of freedom are equal (figure\,\ref{fig:states} right column). This thermal equilibrium state corresponds to the situation of two independently created quasi-condensates where the phase profiles of the clouds are uncorrelated down to the thermal phase correlation lengths, resulting in a relative phase profile that fluctuates strongly along the length of the system\,\cite{Hofferberth2008,Betz2011}. In experiment, the thermal equilibrium situation can be purposely created, by splitting a thermal gas into two, followed by the creation of two independent quasi-condensates through further cooling in the double-well.

The transition between the initial non-equilibrium state and the thermal equilibrium state can therefore be directly probed by studying the relative phase fluctuations. 

Considering above, we note that any thermalization mechanism must necessarily redistribute the mode populations of the relative and common degrees of freedom to lead to the thermal equilibrium situation.
 
%

\subsection{Full Distribution Functions of Matter Wave Interference Contrast}
Previous experiments have also investigated the dynamics of split Bose gases~\cite{Hofferberth2007a,Jo2007} or equivalent systems~\cite{Widera2008}. In these works, though, due to the lack of theoretical or experimental tools for a full characterization of the transient states, the driving mechanism of the dynamics could not or only indirectly be revealed.  Consequently, the state to which the system decayed remained elusive. 

This directly illustrates the key challenge common for most experiments trying to observe non-equilibrium dynamics: the scarcity of experimental tools for characterizing the complex many-body states during the evolution. In this work, this problem is approached by generalizing tools developed for studying equilibrium systems of spatially separated 1d Bose gases~\cite{Polkovnikov2006,Hofferberth2008,Stimming2010,Betz2011} by applying them to the non-equilibrium case, as recently proposed in~\cite{Kitagawa2010,Kitagawa2011}. For this purpose, we measure the time evolution of \textit{full quantum mechanical probability distribution functions (FDFs)}. 

When dealing with many-body quantum systems, valuable information about the underlying quantum state of the system can be gained through noise correlation measurements. 
Performing such measurements already deepened our understanding of quantum mechanics as it led to the discovery of the Hanbury Brown-Twiss effect\,\cite{BROWN1956} which triggered the development of modern quantum optics\,\cite{Glauber1963}. Furthermore,  the study of current fluctuations led 
to important observations in quantum-Hall systems\,\cite{Reznikov1997,Saminadayar1997}. Recently in atomic physics the analysis of noise correlations revealed the coherence properties of atom lasers\,\cite{Ottl2005} and enabled observations of the Hanbury Brown-Twiss effect for massive fermions and bosons~\cite{Aspect2008}. It was further suggested\,\cite{Altman2004} and experimentally demonstrated\,\cite{Folling2005,Spielman2007,Rom2006,Guarrera2008,Perrin2012} that noise correlations in time-of-flight can be used to probe strongly-correlated equilibrium states of many-body quantum systems.

In 1d systems fluctuations play a much more pronounced role than in their three dimensional (3d) counterparts~\cite{Mermin1966,Hohenberg1967}. The interference pattern of two expanding 1d quasi-condensates therefore inherently contains strong noise and fluctuations that can be used to study the many-body state of the system. When performing interference experiments with spatially separated 1d Bose gases, one is furthermore in the advantageous position that the interference pattern itself is directly related to the correlation functions of the system. For example, the mean squared value of the integrated matter wave interference contrast $\braketone{|{C}|^2}$ is a measure of the integrated two-point phase correlation function. Similarly, by exploiting the strong shot to shot fluctuations of the contrast, higher moments of the contrast can be obtained which contain information about higher-order correlation functions~\cite{Polkovnikov2006,Gritsev2006}. An experimentally accessible quantity capturing all these higher moments of the contrast and therefore also the higher-order correlation functions is given by the full distribution function of the contrast $P(C)$. This function describes all the fluctuations of $C$ in the measurement, as $P(C)\,dC$ measures the probability of observing a contrast between $C$ and $C+dC$. Higher moments of $\braketone{C^k}$ of the contrast can be obtained by integration $\int{C^k P(C)\,dC}$. 

It is important to point out that the method of using FDFs to characterize a system requires the detection of \textit{single} realizations of the quantum system in question. If only ensemble averages can be measured in the experiment, the statistics of those values will always be Gaussian due to the central limit theorem and the characteristic higher moments of the observable will not be accessible. The method of analyzing quantum states through the FDFs has already been successfully used to characterize isolated~\cite{Hofferberth2008} and tunnel-coupled~\cite{Betz2011} systems of two separated 1d Bose gases in equilibrium. Within the context of~\cite{Kitagawa2010, Kitagawa2011}, this method was extended theoretically to non-equilibrium systems, whereas the experimental demonstration of this method is the topic of the work presented here and in~\cite{Gring2012}. 
\section{Experimental Details} \label{Experimental details}
\begin{figure}[tb]
\begin{center}
\includegraphics[width = 0.9\textwidth]{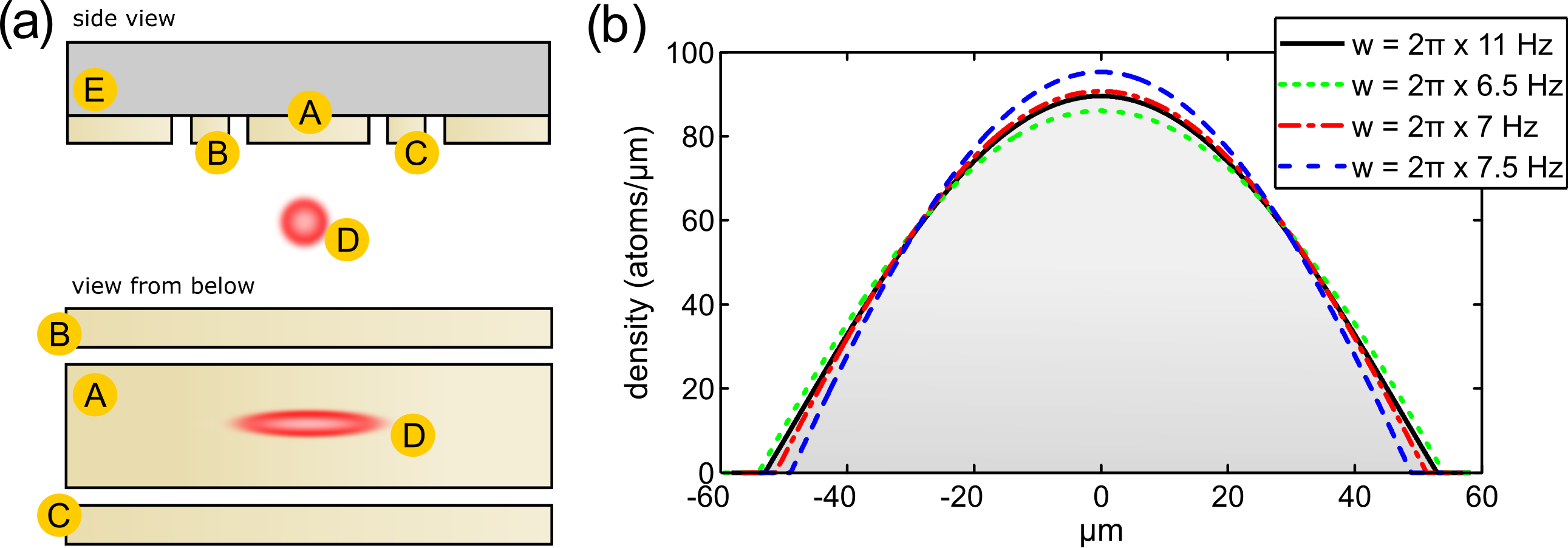}
\caption{(a) Schematic of the atom chip layout. (A) The $100\,\mu$m-wide central trapping wire, (B,C) $30\,\mu$m-wide wires for RF currents, (D) the atom cloud, (E) the silicon substrate.  Additional structures, at a distance of $\sim 1\,$mm from the condensate, are used to provide longitudinal confinement. 
(b) Simulation of the longitudinal density profiles of trapped quasi-condensates gases following~\cite{Gerbier2004}. The black solid line shows the profile of the cloud before splitting for an atom number of 6000 atoms. The colored dotted, dotted dashed and dashed lines show the profile of the sum density of two quasi-condensates trapped in the double-well trap with a longitudinal trapping frequency of 6.5, 7, and 7.5\,Hz. For the experimentally realized double-well trapping frequency of 7\,Hz, the profile matches almost exactly to that of the single-well trap at 11\,Hz. 
}
\label{fig:ChipSetup}
\end{center}
\end{figure}

The preparation of the atomic sample is performed following our standard procedure~\cite{Wildermuth2004a} using magnetically trapped \rb in the $5S_{1/2}\: F=2, m_F=2$ state on an atom chip~\cite{AtomChips2011}. The atom chip is a current-carrying gold structure micro-fabricated on a silicon substrate. A basic outline of the most important structures on the chip is given in figure \,\ref{fig:ChipSetup}a. 

\subsection{Realizing a Quench: The Coherent Splitting Process} \label{Creation of the double-well system}

We first create a single-well trap using the static magnetic field of a 100\,\um wide wire in combination with homogeneous bias fields. Longitudinal confinement is provided by additional wires. The trap is located at a distance of about 100\,\um away from the atom chip surface (see figure\,\ref{fig:ChipSetup}a).  This strongly anisotropic trap has measured trap frequencies of $\omega_\perp = 2\pi \times (2.1\pm0.1)$~kHz (radial confinement) and $\omega_\parallel = 2\pi \times (11\pm1)$~Hz (axial confinement). 

To create the double-well potential we use an adiabatic dressed-state potential~\cite{Hofferberth2006,Lesanovsky2006}. By applying RF current to two 30\,$\mu$m-wide wires that are adjacent to the central 100\,\um wire, the cloud is split along the radial direction of the trap and perpendicular to the direction of gravity. This ensures that any potential sag due to gravity is common for both wells and that the $1/r$ dependence of the RF fields emanating from the wires is the same for both wells. For the splitting process, the amplitude of the RF current is linearly increased from 0 to typically 22.5\,mA within 17\,ms and the frequency is 30\,kHz detuned to the red of the $F=2, m_F=2 \rightarrow F=2,m_F=1$ transition at the minimum of the initial static magnetic trap. This creates a double-well with a separation of $(2.75 \pm 0.05)\,\mu$m, a simulated barrier height of $(2.9\pm0.1)\,$kHz and measured trap frequencies for each well of {$\omega_\perp = 2 \pi \times (1.4\pm0.1)\,$kHz} and $\omega_\parallel=2 \pi \times (7 \pm 1)\,$Hz. Great care was taken to guarantee stability and repeatability of the splitting process, especially to gain a symmetric splitting with almost zero mean atom number difference between the two halves. For the fine tuning of this quantity the relative amplitude of the RF current applied to the two 30\,$\mu$m-wide wires can be slightly adjusted. 

Following~\cite{Betz2011,Ananikian2006}, any residual tunnel coupling between the two halves of the system was calculated to be $J \ll 2\pi \times 0.1\,$Hz. In this regime, a residual tunnel coupling would have no significant effect on the results presented in this work~\cite{Stimming2010}. The absence of tunnel coupling was also confirmed by experiments with independently created condensates in the same double-well~\cite{Betz2011}. In the measurements presented in section \ref{Longterm} we further used a larger final RF amplitude which resulted in a larger splitting distance and larger barrier height corresponding to an even lower bound for the residual coupling and found no difference to the previous results.

Although the splitting RF current was ramped up in a time of typically 17\,ms, the actual splitting process is much faster than the timescale of the longitudinal dynamics and happens close to the end of the ramp when the coupling between the two gases vanishes. From the evolution of the mean relative phase between the two gases for deliberately imbalanced double-wells~\cite{Schumm2005}, we determined this decoupling to occur $(15\pm0.5)\,$ms after the start of the RF ramp. All evolution times given in the following refer to this point in time. To capture also the evolution that happens after the splitting but still during the ramp up, it is possible to probe the atoms during the ramp up.

While the actual splitting is much faster than the ramp up of the RF current, the trap is nevertheless continuously deformed throughout the whole ramp up time. The choice of 17\,ms total duration is a compromise between realizing a fast splitting and avoiding the excitation of collective oscillations of the cloud. The trapping frequencies of the final double-well were chosen to be such that the longitudinal profile of the clouds before and after splitting are almost exactly matched, as can be seen from the simulations in figure\,\ref{fig:ChipSetup}. This minimizes the creation of longitudinal breathing oscillations of the clouds during splitting. As we will discuss in section~\ref{dephasing} these breathing oscillations and the general instability of the splitting process severely limited previous studies of coherently split 1d Bose gases~\cite{Hofferberth2007a}. We find that longitudinal center-of-mass oscillations of the clouds after splitting can not be completely avoided without the use of more complex splitting protocols. However, these small oscillations do not significantly alter the physics as they do not change any significant parameters of the cloud such as density or total size.


\subsection{Probing the System}

\begin{figure}[ht]
\centering
\includegraphics[width=0.6\textwidth]{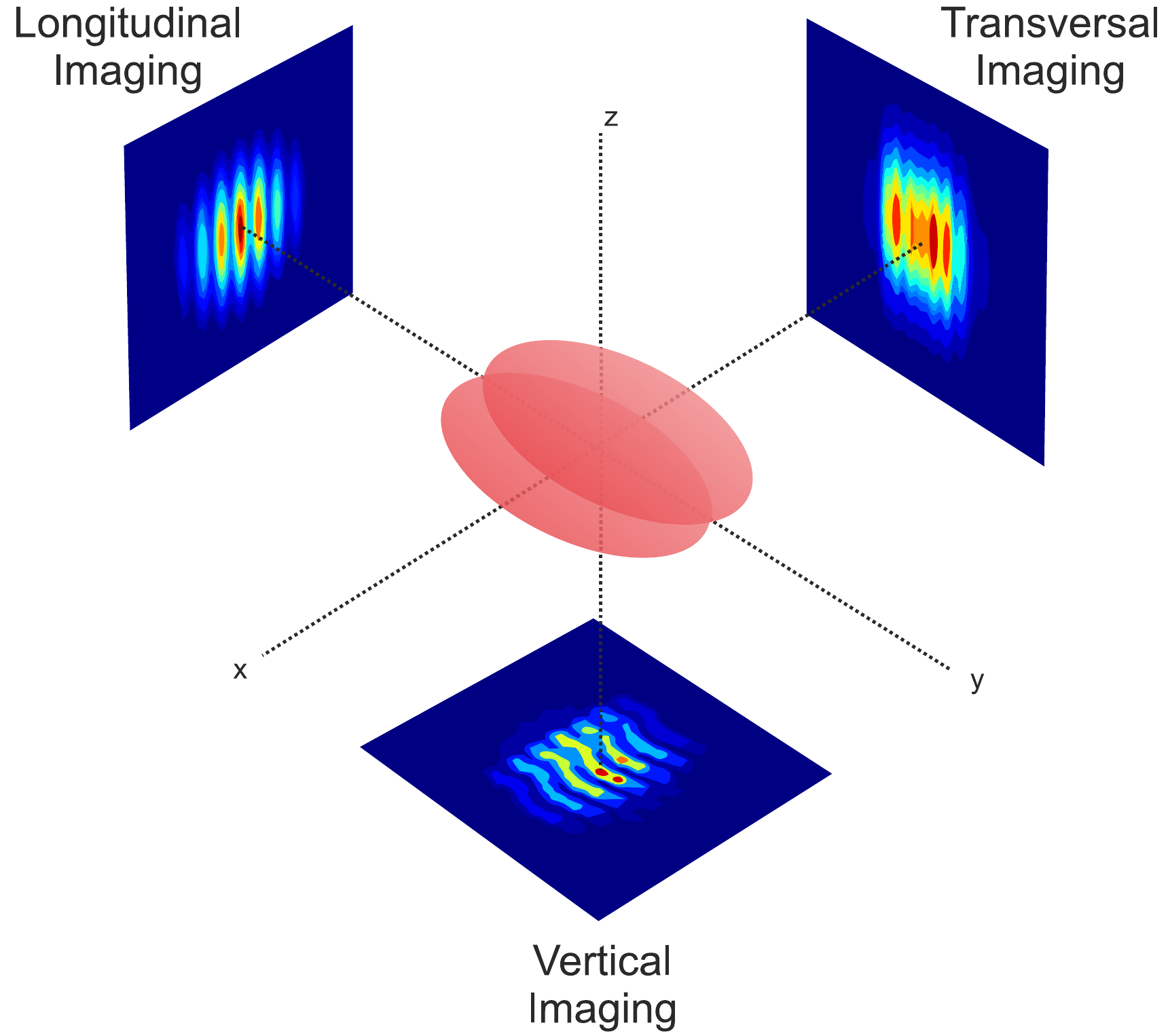}
\caption{Schematic view of the experimental situation considered. Splitting of the gases is performed along the $x$-axis, the $y$-axis corresponds to the longitudinal direction of the 1d Bose gases, gravity points downwards along the $z$-axis. The atomic density distribution in time-of-flight can be imaged along each axis of the coordinate system using absorption imaging. Imaging along the $x$-axis is used to transversally probe single gases before the splitting or the incoherent sum of the two independent gases in the double well, the latter case being depicted in the figure. The corresponding absorption picture shows strong density fluctuations which are used for thermometry (details see text). Imaging along the $y$-axis and the $z$-axis is used to probe the interference pattern of two 1d Bose gases. Imaging along the $y$-axis results in a projection of the interference pattern along its length, while imaging along the $z$-axis reveals the full undulating structure of the interference pattern.}
\label{interfering clouds}
\end{figure}

The system is studied using absorption imaging, which integrates the 3d structure of the expanding atomic density distribution along one direction of space. Figure \ref{interfering clouds} shows schematic views of the integrated density distribution as seen from each of the three detection directions realized in the experiment. 

Imaging along the $x$-direction is used to probe the system from its transversal direction. This can be applied both to extract information about a single quasi-condensate before the splitting process as well as about the incoherent sum of two quasi-condensates expanding from the double-well. As the phase of a quasi-condensate fluctuates in space, density fluctuations will appear along its length when it is released from the trap and allowed to expand in time-of-flight\,\cite{Dettmer2001,Imambekov2009,Manz2010a}. 
These spatial density fluctuations are a direct consequence of matter wave interference between the different positions along the single quasi-condensate which have different phases. In equilibrium, the in-situ phase profiles are determined by the temperature of the system, so that the measurement of the spectrum of density fluctuations in time of flight can be used for thermometry of the system before the splitting process, as experimentally demonstrated in~\cite{Manz2010a}. All initial temperatures mentioned in the following, as well as the temperatures of pairs of independently created gases in the double-well have been determined using this method. 

Information about the relative phase between the two 1d clouds can be obtained from absorption images along the longitudinal axis of the two gases in the $y$-direction, or transversally along the vertical $z$-axis. 

In an intuitive picture, the interference pattern can be thought of as an array of many thin interference patterns stacked up together along the longitudinal direction of the cloud, where the phase of each thin interference pattern depends on the local in situ phase difference between the clouds. The axially integrated contrast $C(L)$ of the interference is therefore a direct measure for the relative phase fluctuations within the integration length $L$~\cite{Bistritzer2007,Polkovnikov2006,Imambekov2008,Schumm2005a}.
In the following we will discuss the technical implementation of the detection systems for the longitudinal and vertical directions in detail and show how we extract the integrated contrast $C(L)$ for different integration lengths $L$. 

\subsubsection{Optical Slicing Method} \label{chapter:slicer}
\begin{figure}[ht]
\centering
\includegraphics[width=\textwidth]{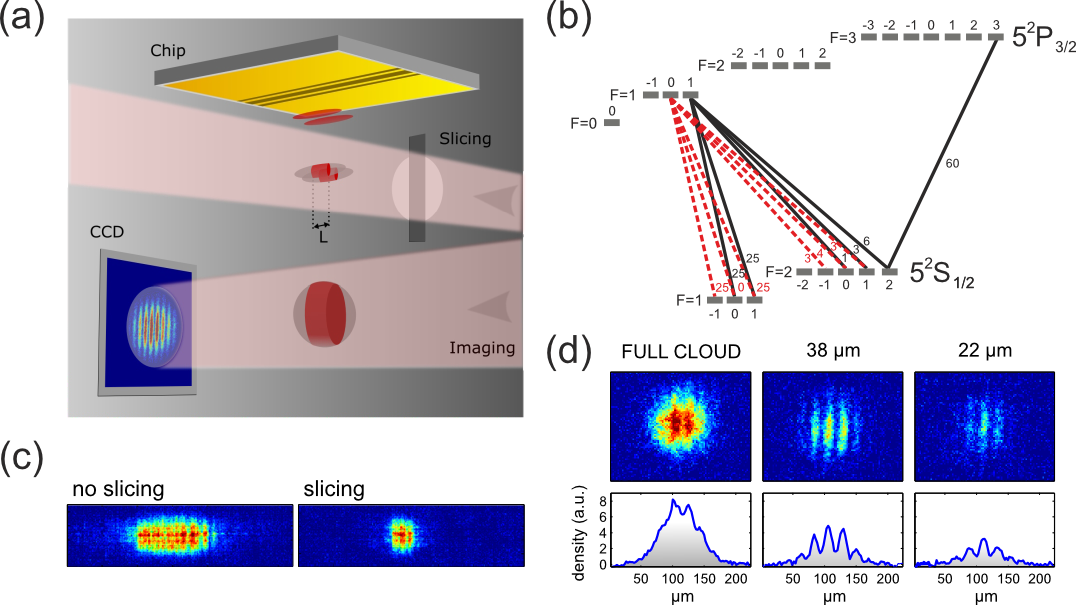}
\caption{(a) Illustration of the working principle of the optical slicing method. The atom clouds are released from their trap and allowed to fall under gravity. Shortly after release, the slicer beam containing a shadow of length $L$ is projected onto the atom clouds to pump the unwanted edges into a state that is dark to the imaging beam light. Standard absorption imaging is then used at a much later time of fight to image the remaining undisturbed part of the cloud. (b) The relevant transitions of the D2 line of $^{87}$Rb for optical pumping. Numbers indicate relative transition strengths using the normalization from~\cite{Metcalf1999}. (c) Images of the effect of slicing on a quasi-condensate as seen from the direction of the slicing beam. (d) Interference images and corresponding line profiles for different integration lengths $L$ after an evolution time of 7\,ms in the double-well potential.}
\label{slicer3D}
\end{figure}

In this scheme, the imaging beam propagates along the longitudinal axis of the 1d gases and directly performs an integration of the interference pattern along the length of the gases. In order to vary the integration length $L$ we perform the following slicing scheme using optical pumping (figure \ref{slicer3D}a). Shortly (1\,ms) after the two halves of the split condensate are released from the trap, a shadow of width $L$ is imaged on the atoms using a $50-125\,\mu$s pulse of optical pumping light. The frequency and the polarization of the light are chosen such that the atoms, initially being in the $5S_{1/2}\: F=2, m_F=2$ state, are excited to a state in the $5P_{3/2} \: F=1$ manifold (figure \ref{slicer3D}b). From there, the atoms preferentially decay to a state in the $5S_{1/2}\: F=1$ manifold which is dark to the imaging beam operating on the $5S_{1/2}\: F=2\: \rightarrow 5P_{3/2}\: F=3$ transition. Atoms which are in the protected shadow area of length $L$ are not optically pumped and can be imaged after whichever chosen time-of-flight using an absorption imaging system operating along the axial direction of the system.

The branching ratios of the $5S_{1/2}\: F=2 \rightarrow 5P_{3/2}\: F=1$ pumping transitions are such that, on average, only very few photons are needed for pumping. Already after one photon has been absorbed and remitted an atom has a probability above 80\% of being in a dark state. This ensures that the atoms remain virtually undisturbed by the optical pumping process. 

To create the sharp-edged shadow for the optical slice, we use a multi-lens high-resolution imaging objective in reverse. The imaging system has a maximum resolution of $2.5\,\mu$m, and a demagnification of approximately 1:4.6. The shadow is produced by a glass target with a series of opaque bars (produced at the ZMNS TU-Wien). In order to be able to change the integration length $L$ in each experiment the bar target is mounted on a computer-controlled motorized translation stage. 

The effect of the optical pumping light on the detected atomic density is presented in figures \ref{slicer3D}c and \ref{slicer3D}d. Figure \ref{slicer3D}c shows examples of sliced atomic density distributions after optical pumping and a subsequent time-of-flight. The images were obtained using the same imaging system ($x$-direction) that is used in reverse for the optical pumping. Example interference images are shown in figure \ref{slicer3D}d for three integration lengths and a hold time of 7\,ms in the double-well potential. Note that due to the evolution in the double-well potential, the contrast of the interference pattern which is integrated over the full cloud is already significantly reduced in most of the images. Furthermore, since for shorter $L$ more and more atoms are cut away, the detected atom number becomes lower, giving a limit to the shortest integration length which can be investigated. 
\subsubsection{Direct Imaging Method} \label{chapter:spotlight}
\begin{figure}[ht]
\centering
\includegraphics[width=0.8\textwidth]{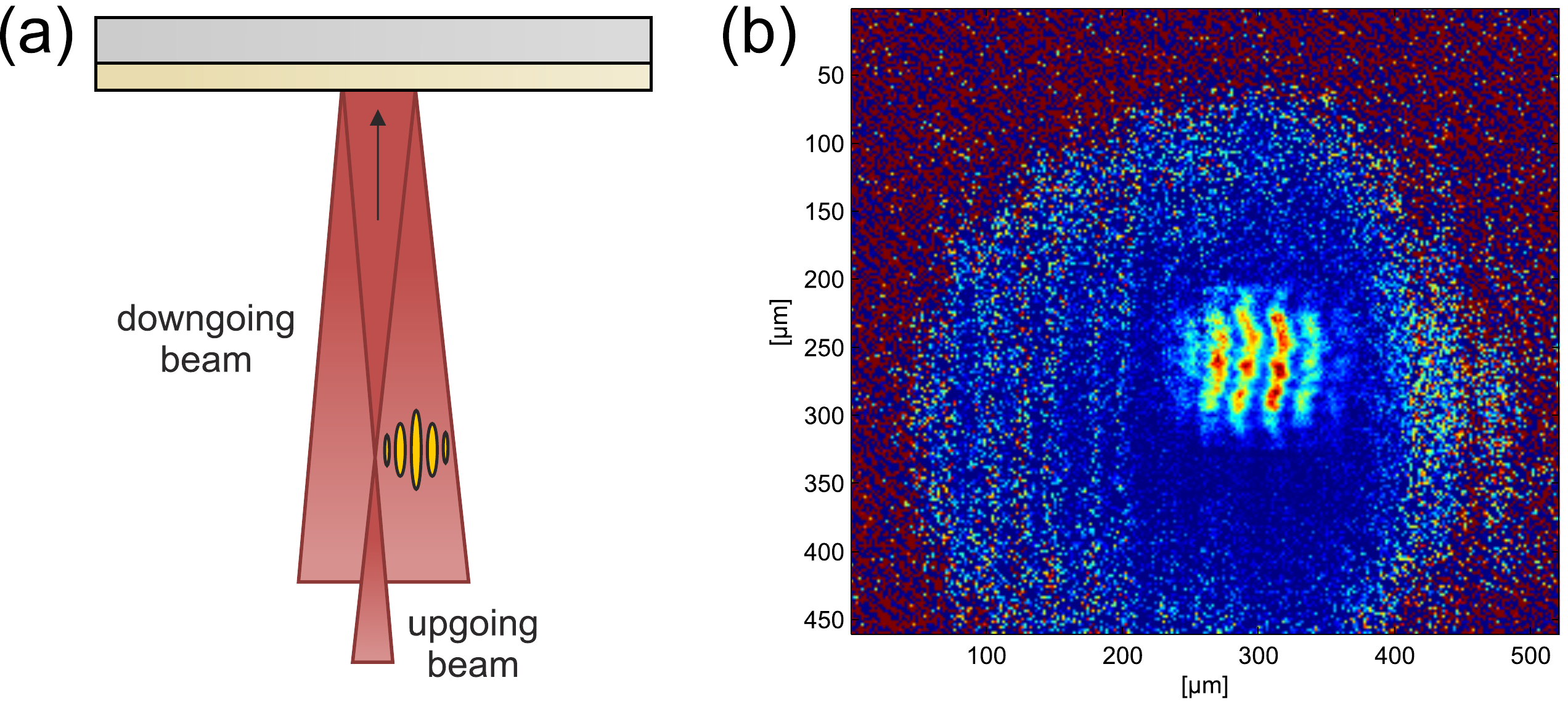}
\caption{(a)  Schematic of the direct imaging method. (b) Exemplary image. As the incident beam is focused, only a very small area around the atoms is illuminated (like a spotlight), whereas the rest of the CCD image is filled with unimportant noise. This removes a second, out of focus image which would be created if the incident beam passed the atoms twice.}
\label{vertical4}
\end{figure}

In the vertical direction we can directly image the undulating structure of the interference pattern and freely select the integration length via post-processing. To this end an imaging system is employed which observes the atoms from below (figure\,\ref{vertical4}).  

The implementation of such an imaging system is challenging since the chip surface blocks the line of sight, hindering standard absorption imaging schemes. One possibility to circumvent this problem is to use a light sheet fluorescence detector as in~\cite{Betz2011,Bucker2009}. Here, we instead rely on absorption imaging, as it allows for a higher imaging resolution. 


To this end, we reflect the imaging beam from the gold surface of the atom chip close to normal incidence. In this procedure the high quality of the gold layer is essential to achieve good images. For a collimated imaging beam the reflection means that the light passes the atom cloud twice~\cite{Smith2011a}. This results in a second, virtual image behind the atom chip, which is out of focus and overlaps with the primary image. Since the imaging of the interference pattern requires the imaging beam to propagate almost parallel to the fringes, it is not possible to offset the two images by using a large angle of incidence when propagating the imaging beam onto the atom chip surface. In order to eliminate the second image we therefore focus the imaging beam close to the atoms, avoiding the first absorption process as shown in figure \ref{vertical4}a. This limits the field of view to the area of a 'spotlight' (figure\,\ref{vertical4}b) and ensures that the light only interacts with the atoms once. Also, by aligning the slowly diverging beam onto the central $100\,\mu$m wire of the chip, interference effects caused by diffraction from the microscopic chip structures can be minimized. We note that a similar method was used in~\cite{Armijo2010}. In figure \,\ref{vertical4}b we show an experimentally obtained image, where the imaging system was set up for a time-of-flight of 16\,ms. The image clearly reveals the full structure of the spatially varying interference pattern, which can then be used for further analysis.

\subsection{Extraction of the Matter Wave Interference Contrast}

The main observable used in this work is the contrast $C(L)$ of the matter wave interference pattern integrated over a length $L$ (figure~\ref{experiment schematics}c). For the optical slicing the integration length $L$ is selected in the imaging setup by choosing a specific mask, and the integration is performed optically. In the direct imaging method the recorded image of the interference pattern is integrated over a the length $L$ during post-processing.  In both cases this results in a line profile (see figure~\ref{experiment schematics}). 
This line profile is fitted with a sinusoidal function $f_L(x)$ having a Gaussian envelope:
\begin{equation}
f_L(x) = e^{-\frac{x^2}{2\omega^2}}  \left[1 + C(L) \cos \left({ \frac{2\pi x}{\Delta}+\theta_L} \right) \right],
\end{equation}
where $\omega$ is the cloud size, $\Delta$ is the fringe spacing and $\theta_L$ is the global phase of the interference pattern integrated over $L$, i.e. the phase of the line profile. As interactions do not play a significant role in the radial time-of-flight expansion for our parameters, the fringe spacing $\Delta$ can be directly related to the double- well-potential separation $d=h t / m \Delta$, where $t$ is the time-of-flight, $m$ is the mass of the atoms and $h$ is Planck's constant~\cite{Andrews1997}. Thus, the stability of the double-well splitting can be monitored with interferometric precision. The contrast $C(L)$ is extracted from this fit and used for further analysis that is presented in the remainder of this paper. Repeated realizations of the experimental cycle then allow us to build the time-dependent FDFs of the interference contrast $C(L)$. 

\subsection{Comparison of the Two Imaging Methods} \label{comparison}

The use of two technically different imaging systems allows us to preclude systematic effects of the imaging system on the observed contrast evolution. 

\begin{figure}[tb]
\centering
\includegraphics[width=0.7\textwidth]{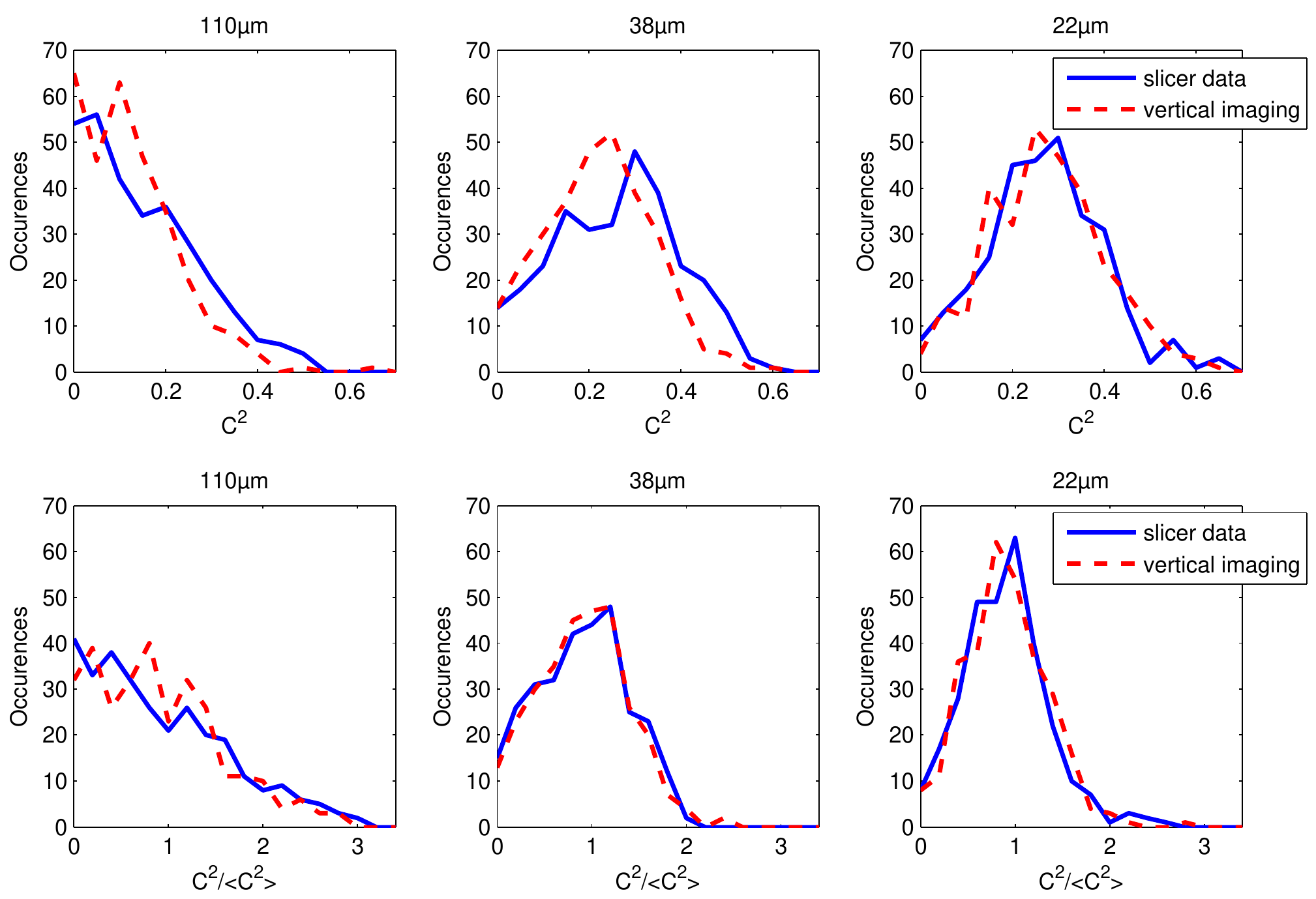}
\caption{Comparison of distributions of the squared contrast obtained with the optical slicing method and the direct imaging system for the same double-well configuration. The evolution time is $10$ms and each distribution contains $300$ realizations, which minimizes statistical effects. In the upper row we plot exemplary distributions of the squared contrast, in the lower row the squared contrast is normalized to its mean. While the former shows slight differences for the results of both imaging systems, the latter shows very good agreement (see text for details). 
}
\label{imcomparison}
\end{figure}

Example FDFs obtained for different integration lengths $L$ from both imaging systems are shown in figure~\ref{imcomparison}. The total area under the curves is normalized to 1 in order to represent a probability density. The upper panel shows FDFs of the squared contrast $C^2$. While the overall form of the FDFs show qualitative agreement, a small offset between the results of the two imaging systems remains. This is due to the fact that the absolute value of $C^2$ is affected by the finite imaging resolution and any mis-alignment of the imaging system. These effects can be eliminated by analyzing contrast distributions that are normalized to the mean contrast $\braketone{C^2}$. After this procedure the offset vanishes, as shown in the lower panel of figure~\ref{imcomparison}. 

This comparison confirms that the form of the  $C^2/\braketone{C^2}$ distributions is not influenced by any technical effects of the imaging systems and can therefore be used to study the dynamics of the system. Note that the normalization to the mean contrast also removes further systematic errors affecting both imaging systems, as for example imperfections in the switch-off of the trapping fields. 

In the following, the only and important practical difference between the two methods is thus that the total measurement time with the direct imaging is significantly shorter than the time needed to obtain the data with the optical slicing method. This is a consequence of the fact that with the direct imaging method only one experimental run is necessary to extract information on all length scales down to the imaging resolution. Systematic studies of large parameters spaces or long evolution times, as presented in section~\ref{Longterm}, are thus always performed with the direct imaging system.

\section{Observation of Prethermalization} \label{Pretherm}
\subsection{Dynamics of the FDFs of Matter Wave Interference Contrast}

\begin{figure}
    \centering
        \includegraphics[width=0.6\textwidth]{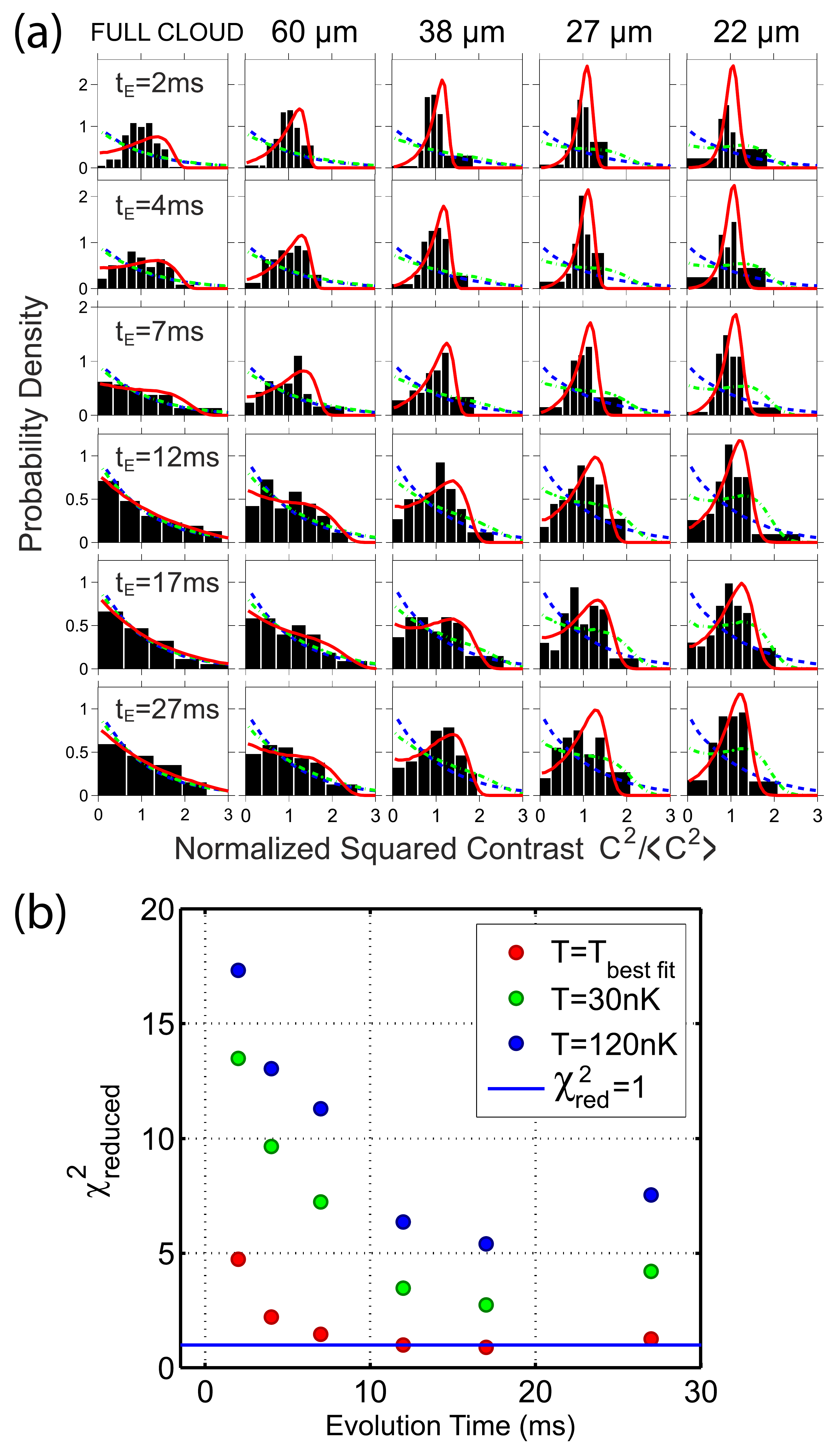}        
    \caption{(a)~Histograms for the distributions of $C^2/\braketone{C^2}$ for evolution times from 2 to 27\,ms after splitting, shown together with the respective best fitting equilibrium distributions, which are $\Teff\equiv T_\mathrm{best\,fit}=(14\pm4), \,(17\pm5),\,(14\pm4)\,$nK for $t_e=12, \,17,\,27 \,$ms, respectively. For comparison, the calculated \textit{equilibrium} distributions for $T=120\,$nK (blue dashed line) and $30\,$nK (green dot-dash line) are added. (b)~The evolution of the best fit's value for the reduced $\chi^2$ with hold time. The probabilities of observing values of the reduced $chi^2$ which are larger than those calculated from the best fits of the measurements are $50\%,\,62\%,\,18\%$ for the last three evolution times of $t_e=12, \,17,\,27 \,$ms respectively.}
    \label{fig:c2equilibriumAllfit}
\end{figure}

Figure \ref{fig:c2equilibriumAllfit}a shows the measured time dependence of the FDFs of interference contrast $C(L)$ for the double-well system which was initialized through a coherent splitting process, as described in the previous section. The data was obtained using the optical slicing detection scheme. The central line density is  $\rho=(32\pm7)$ atoms/$\mu$m (error given by standard deviation) in each well at an initial temperature of the unsplit cloud of $T=(120\pm30)\,$nK. For the initial gas before the splitting this results in a chemical potential of $\mu \sim (2/3)\hbar \omega_r$ and an energy $k_B T \sim \,\hbar \omega_r$ due to temperature. These parameters are similar to the parameters of other experiments with 1d gases in microtraps, e.g.~\cite{Amerongen2008,Armijo2011}, and give an effectively 1d Bose gas in the quasi-condensate regime.


After the splitting, a slight net mean imbalance $\Delta n=(N_L-N_R)/(N_L+N_R)$ of $\Delta n=(1.5\pm1)$\% is present. Here, $N_L$ and $N_R$ are the atom numbers of the left and right well, respectively. Due to technical noise the width of this imbalance distribution is around $2-3$ times larger than the quantum shot noise limit of $\sqrt{N}$, where $N=N_L+N_R$ is the total atom number of the system. The imbalance has mainly two effects. The net difference in chemical potential resulting from the mean imbalance leads to a trivial common phase evolution of the two wells\,\cite{Schumm2005a,Lewenstein1996,Javanainen1999}. In addition to this, the width of the imbalance distribution leads to an additional broadening of the global phase distribution. It is a strength of our approach that this technical noise does not affect the dynamics of the contrast distributions, as its effect on the dynamics is negligible compared to the quantum noise associated with the splitting process for the probed integration lengths $L$.

For the shortest evolution time $t_e=2$\,ms the $C^2/\braketone{C^2}$ distributions are peaked on all lengths $L$, being almost all identical in form and having a low probability of obtaining a low contrast $C(L)$. As time evolves, the distributions behave very differently on different length scales. For long integration lengths $L$, the FDFs evolve to become exponential in form, i.e. there is a high probability of observing a low contrast $C(L)$ and the initial correlations of the system appear to be lost. For short integration lengths, however, the FDFs remain peaked and there is a very low probability of observing a low contrast $C(L)$. This directly demonstrates the persisting memory in the system of the initial state. Furthermore, the strong length dependence is a direct signature of the multimode nature of 1d Bose gases~\cite{Widera2008, Kuhnert2012}. Finally, the shape of the distributions evolves very quickly for the first $12\,$ms and very little between $t_e=12-27$\,ms, showing that the system reaches a quasi-steady state. This imposes the question of whether this state corresponds to the thermal equilibrium state of the system. 

\subsection{Emergence of an Effective Temperature}

We will now address the question of whether the observed quasi-steady state already corresponds to the thermal equilibrium state of the system. 

If the system has reached thermal equilibrium, the measured distributions plotted in figure \ref{fig:c2equilibriumAllfit}a should be consistent with equilibrium theory\,\cite{Polkovnikov2006,Gritsev2006,Stimming2010}. 
The obtained \textit{non-equilibrium} distributions of $C^2/\langle C^2\rangle$ were therefore fitted with \textit{equilibrium} distributions taking temperature as a free fit parameter. To this end, a $\chi^2$ analysis~\cite{Hughes2010} was performed. 
Figure \ref{fig:c2equilibriumAllfit}a shows the experimentally obtained distributions together with the respective best fitting equilibrium curves. The agreement is very good for longer evolution times, but less good for early evolution times.  

Figure \ref{fig:c2equilibriumAllfit}b shows the corresponding reduced $\chi^2$ values of the fits. For short evolution times the reduced $\chi^2$ differ significantly from 1, showing that the experimentally obtained distributions are inconsistent with equilibrium theory. 
For evolution times longer than 12\,ms, the observed value of the reduced $\chi^2$ approaches 1, which shows that the experimental data agrees well with equilibrium theory. This indicates that the thermal-like nature of the distributions is established dynamically during the evolution. 

It is thus possible to associate the observed steady state with a temperature $\Teff$ extracted from equilibrium theory. For the last three evolution times $t_e=12, \,17,\,27 \,$ms the fits give $T_{\mathrm{eff}}=(14\pm4), \,(17\pm5),\,(14\pm4)\,$nK, respectively. This is, however,  a factor of \textit{eight} lower than the initial temperature of the unsplit system. Furthermore, the corresponding effective thermal phase coherence length of the relative phase between the two gases is $\lambda_{\phi} = \hbar^2\rho/m k_B T_\mathrm{eff}= 13^{+5}_{-3},\,10^{+5}_{-2},\,13^{+5}_{-3}\,\mu$m. This is much longer than the expected $\lambda_{T}=(1.5\pm0.6)\,\mu$m deduced from the initial temperature of the unsplit cloud. For a detailed study of the emergence and properties of this length scale see~\cite{Kuhnert2012}. 

\begin{figure}[tbh]
    \centering
        \includegraphics[width=0.6\textwidth]{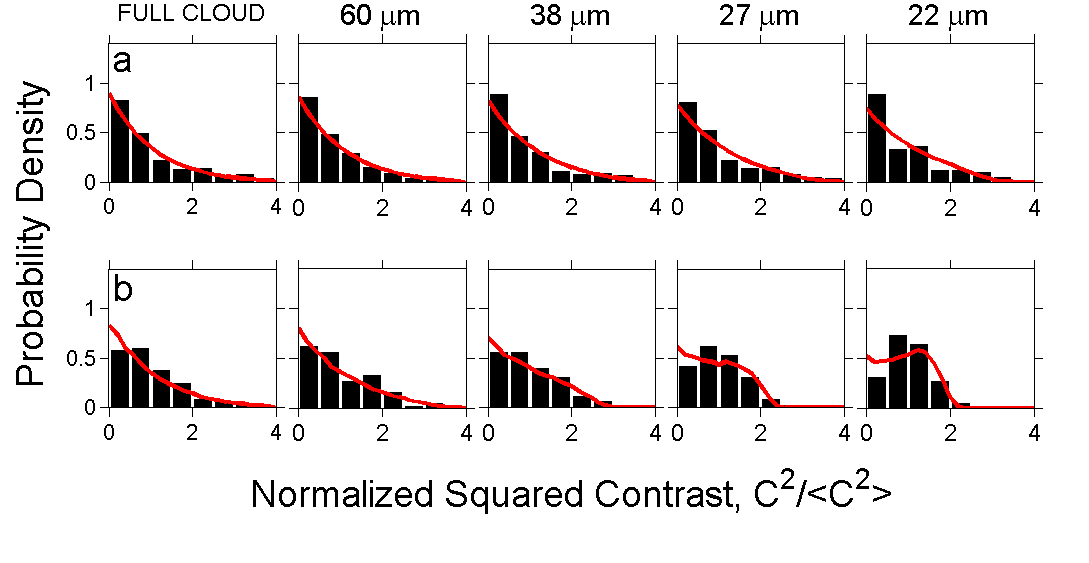}
    \caption{Experimental (histograms) and theoretical (red solid lines) \textit{equilibrium} distributions. (a) $T=(60\pm20)\,$nK, density of $(32\pm6)\,$atoms/$\mu$m. Each plot contains the statistics of $\sim170$ measurements.
    (b) $T=(27\pm7)\,$nK, a density of $(27\pm5)\,$atoms/$\mu$m, statistics of $\sim90$ measurements. Both datasets show very good agreement with the theoretical description.}
    \label{fig:c2equilibrium}
\end{figure}

As a further illustration of the striking difference between the observed FDFs and the ones expected for the corresponding equilibrium system, we performed interference experiments with two independently created 1d gases, employing the same double-well potential as for the non-equilibrium situation. The measured equilibrium distributions are plotted in figure \ref{fig:c2equilibrium}. Even for $T=(60\pm20)\,$nK the measured $C^2/\braketone{C^2}$  distributions are exponential in form for all length scales probed, which is in stark contrast to the observed non-equilibrium steady state distributions but in perfect agreement with what is expected from equilibrium theory. A second, colder dataset at $T=(27\pm7)\,$nK visualizes the crossover from an exponential to a Gumbel-like shape of the equilibrium distributions~\cite{Hofferberth2008,Gritsev2006}. 

It is, however, a justified question as to whether the very low observed $\Teff$ of the quasi-steady state could also be explained by a cooling effect involved in the splitting process or, stated differently, how does the initial temperature of the unsplit cloud relate to the final temperature of the gas after splitting and after a possible equilibration? In principle the rapid splitting procedure is a rather violent process which is very likely to transfer some energy to the system. This would suggest that the final equilibrium temperature should be even higher than the initial temperature of the unsplit system, in contrast to the steady-state that we observe.  

On the other hand, the splitting process leads to a decompression of the gas, which, for a 3d ideal gas would lead to a temperature reduction. If we assume an adiabatic splitting 
a comparison of the trap frequencies leads to a lower bound on the decompressed temperature of $(0.6\pm0.1) \times T$~\cite{Ketterle1999}. This temperature reduction has been confirmed by preparing a thermal 3d cloud of atoms in the initial single-well trap and turning on the double-well potential using the same 17\,ms ramp as in the non-equilibrium measurements. For this procedure, we observe a decrease in temperature of $(0.59\pm0.15)\times T$ in agreement with decompression. 

To our knowledge, no quantitative descriptions of this process exist for degenerate 1d gases. The exact evolution and distribution of the thermal energy during the splitting therefore remains a topic of ongoing research. 
The only quantities one can compare to are the initial temperature $T=(120\pm30)$~nK of the unsplit system and the lower bound of $T \sim (70\pm20)\,$nK calculated under the assumption of an adiabatic decompression. Figure \ref{fig:c2equilibriumAllfit} therefore also shows the obtained values for the reduced $\chi^2$ if the experimentally obtained non-equilibrium distributions are fitted with equilibrium theory for $T=120$\,nK. The plot shows that these high temperatures are clearly rejected by the $\chi^2$ test. In addition, in figure \ref{fig:c2equilibriumAllfit} the corresponding equilibrium distributions for $T=30\,$nK (green line) are plotted for comparison. This shows that even for $T=30\,$nK, which is much lower than the lower temperature bound obtained from decompression, the equilibrium distributions are still significantly different from what is observed in the experiment for the non-equilibrium system. 

These observations directly demonstrate that the quasi-steady state we observe has thermal-like properties but is not the true thermal equilibrium of the system. We thus associate this quasi-steady state with prethermalization, as introduced in~\cite{Berges2004} and suggested for split 1d Bose gases in~\cite{Kitagawa2011}. The system decays rapidly to a state whose contrast distributions are thermal-like in form, exhibiting a temperature almost an order of magnitude lower than the temperature of the initial unsplit system. As we will see in section\,\ref{LL Description}, the relaxation of the system to this prethermalized state is very well described by a dephasing of the momentum modes of the system~\cite{Kitagawa2010,Bistritzer2007, Kuhnert2012}. We will first briefly discuss the persistence of this prethermalized state as the system continues to evolve.

\section{Long-term Evolution of the System} \label{Longterm}

As shown in section \ref{Pretherm}, the non-equilibrium system decays rapidly to a prethermalized state. However, the question still remains as to any subsequent evolution of the system, particularly whether the system will eventually reach the true thermal equilibrium corresponding to two independent gases.

In the measurements presented in the following, the initial temperature of the unsplit system was $(78\pm10) $\,nK and the peak atomic line density of the split system was $(38\pm9)$\,atoms$/\mu$m in each of the two wells, i.e. approximately similar parameters to those presented in section\,\ref{Pretherm} above. Furthermore, the final amplitude of the current for the dressing RF creating the double-well was increased to 23.8\,mA leading to a higher potential barrier, larger separation of the double-well and therefore placing an even lower upper bound on any hypothetical residual tunnel-coupling. To ensure that the actual splitting process was also the same, the same linear ramp speed as in the measurements in section\,\ref{Pretherm} was used, but the total ramp duration was extended to 18\,ms. 

\begin{figure}[tbh]
    \centering
        \includegraphics[width=0.6\textwidth]{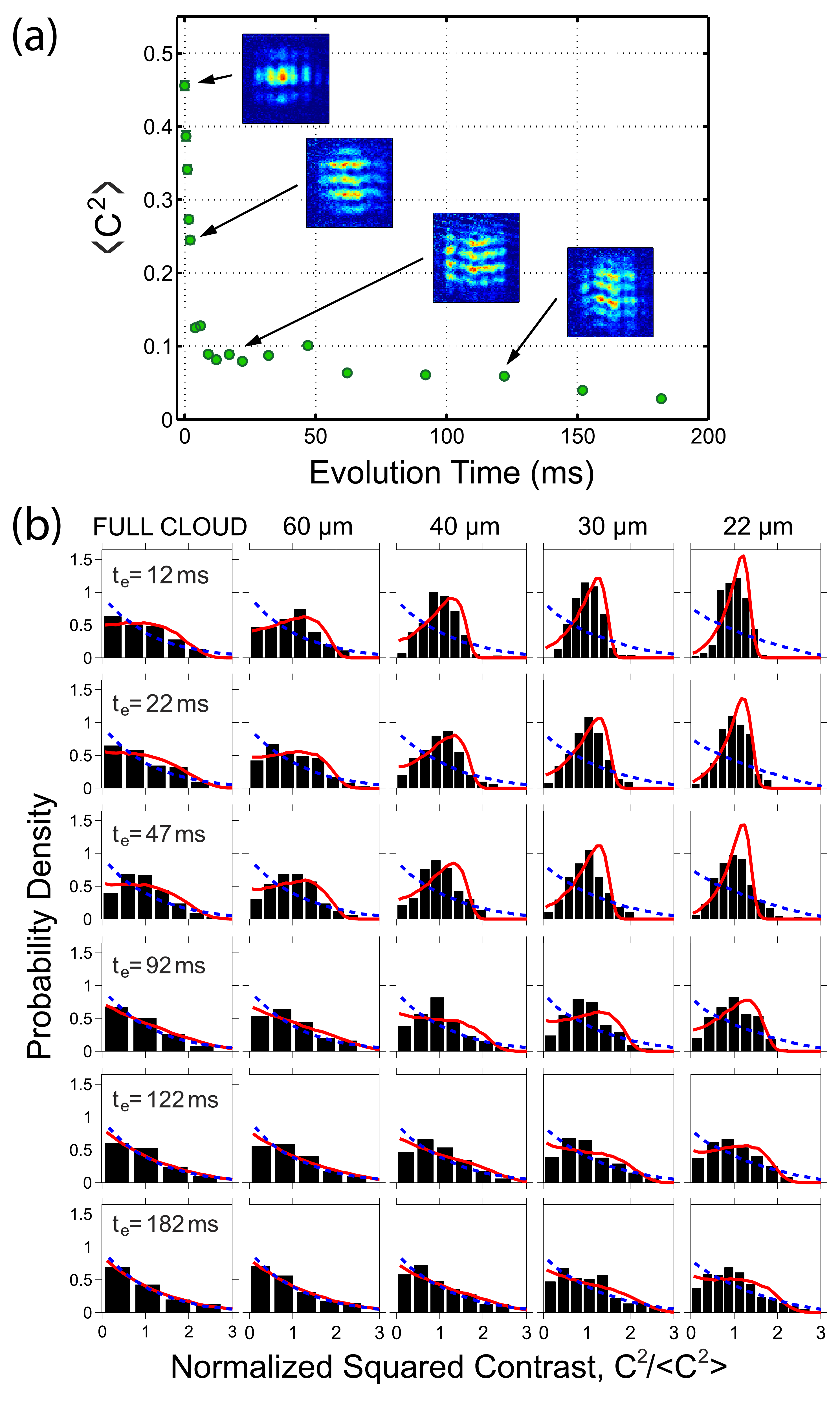}
        
    \caption{(a) Time dependence of the mean squared contrast $\braketone{C^2}$\,\cite{Gring2012, Kuhnert2012} together with example pictures obtained using the direct imaging system. 
    (b) FDFs of $C^2/\braketone{C^2}$ extending to longer evolution times. Again, the experimentally obtained data (histograms) can be very well described by equilibrium theory where the distributions obtained at the best fitting temperature are plotted using red solid lines. The blue dashed lines show theoretical equilibrium distributions for the initial temperature $T=78\pm10$\,nK of the unsplit cloud.}
    \label{scan687_I}
\end{figure}

\begin{figure}[ht]
    \centering
        \includegraphics[width=\textwidth]{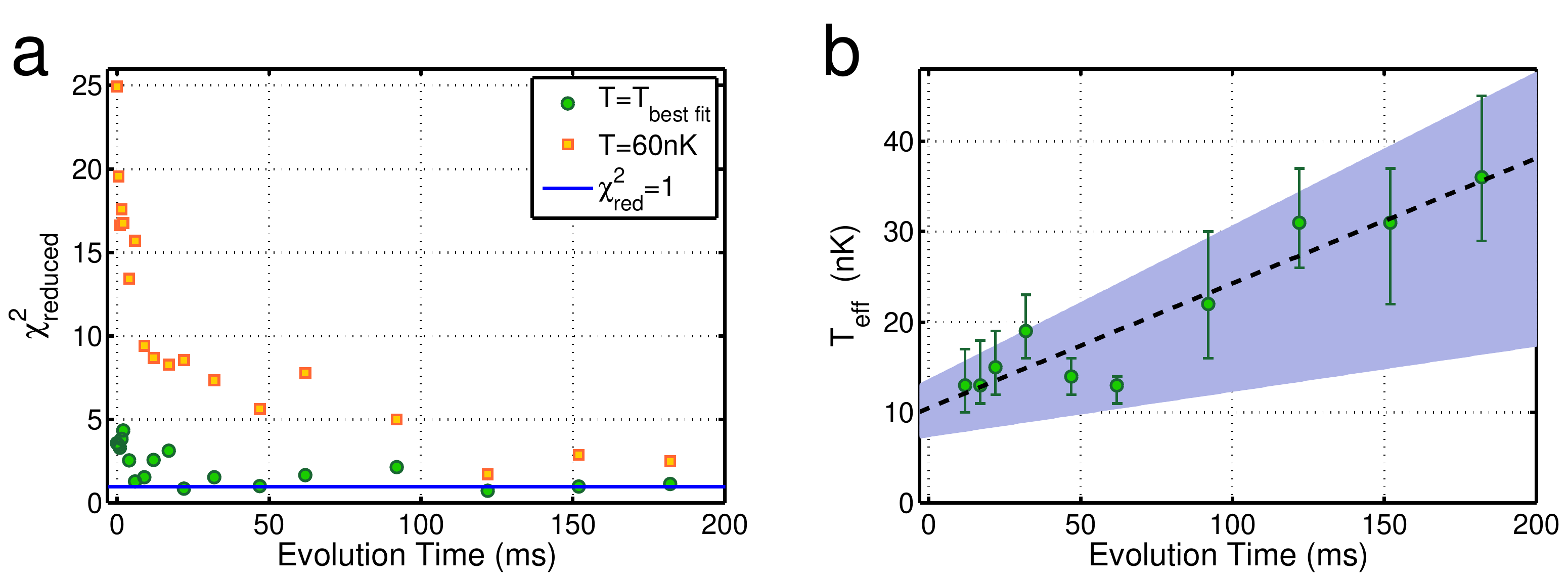}        
\caption{(a) Evolution of the reduced $\chi^2$ with hold time in the double-well potential. (b) Evolution of $\Teff$ for the whole data set extracted using fits to theoretical equilibrium distributions (figure from~\cite{Gring2012}). The dashed linear fit indicates an increase of $\Teff$ over time of $(0.14\pm0.04)$\,nK/ms. The colored area indicates the measured heating rate of the atom trap of $0.11\pm0.06$\,nK/ms, with the area being bounded by the error.}
    \label{scan687_II}
\end{figure}

Figures \ref{scan687_I} and \ref{scan687_II} summarize the main results of the measurements. Figure \ref{scan687_I}a shows the evolution up to a time of almost 200\,ms of the mean squared contrast $\braketone{C^2}$ for the full length of the system. The fast initial decay to the prethermalized state, followed by a very slow further evolution, is clearly visible. As an inset, example interference pictures obtained using the direct imaging system are shown. For very short evolution times the fringes are almost straight, directly visualizing the coherence of the splitting process. For larger evolution times the fringes become more and more wiggly, leading to the observed strong decrease in the integrated interference contrast. Again, for evolution times after the initial decay, the distributions of  $C^2/\braketone{C^2}$ can be described by equilibrium theory, as shown for example in figure\,\ref{scan687_I}b for various evolution times $t_e=12-182$\,ms. 

Figure \ref{scan687_II}a displays the minimal reduced $\chi^2$ values for the inspected evolution times 
which shows that for evolution times after the initial rapid decay, the reduced $\chi^2$ is similar to 1. 
The extracted effective temperatures are again very low compared to the initial temperature of the system, while higher temperatures clearly fail to describe the measured distributions, as demonstrated by the $\chi^2$ analysis for larger temperatures in figure\,\ref{scan687_II}a. 

Observing the long slow decay of $\braketone{C^2}$ seen in figure\,\ref{scan687_II}a 
one may be tempted to assume that this is the system slowly thermalizing. In order to analyze this slow subsequent evolution of the quasi-steady state, the effective temperatures for all times after the initial decay were extracted, as shown in figure \ref{scan687_II}b. As can be seen, $\Teff$ rises slowly over time at a rate of $(0.14\pm0.04)$\,nK/ms. This is, however, comparable to the measured heating rate of the atom trap of $(0.11\pm0.06)$\,nK/ms which was characterized independently using equilibrium quasi-condensates. This indicates that either no thermalization is present, or, if it is present, that it is a very subtle process.  

\section{Luttinger Liquid Description of the Dephasing of the System} \label{LL Description}

In the last sections it was shown that the system created through the coherent splitting of a single quasi-condensate rapidly relaxes to a quasi-steady state that has thermal characteristics but which does not correspond to the true thermal equilibrium state of the system. This observation can be very well understood by the theoretical model presented in~\cite{Kitagawa2010,Kitagawa2011} where the appearance of a thermal-like state is explained through the dephasing of the multimode system. 
Before presenting a quantitative comparison of this model to our experiment we summarize the most important points of this theory in order to give an intuitive picture for an understanding of the prethermalized state. For a more detailed discussion the reader is referred to the original work~\cite{Kitagawa2010,Kitagawa2011}. For an alternative description of the rapidly split quasi-condensate in terms of an effective-tunneling model, see~\cite{Langen2012}.

\subsection{Summary of the Theoretical Model} \label{Summary of the theoretic model}

The evolution of the interference pattern is directly determined by the evolution of the relative phase between the two uncoupled halves of the split gas. We describe this evolution using a Tomonaga-Luttinger liquid approach. In this low-energy approximation, the relative degrees of freedom $\hat{ \phi}(y)$ and $ \hat{ n}(y)$ perfectly decouple from the common degrees of freedom of the system $\hat{\phi}_\mathrm{com}(y)$ and $\hat{ n}_\mathrm{com}(y)$, where the experimental observables are now described by operators. The resulting Hamiltonian for the relative degrees of freedom is of the form $\hat{H} = \frac{\hbar c}{2}\int_{-L/2}^{L/2} dy \left[ \frac{K}{\pi}(\nabla \hat{\phi}(y))^2 + \frac{\pi}{K}\hat{n}^2(y)\right]$, where $K=\pi\xi_h\rho$ is the Luttinger parameter, $c$ is the speed of sound, $\xi_h$ is the healing length and $\rho$ is the atomic density.  
 
\begin{figure}[tb]
\centering
\includegraphics[width=0.7\textwidth]{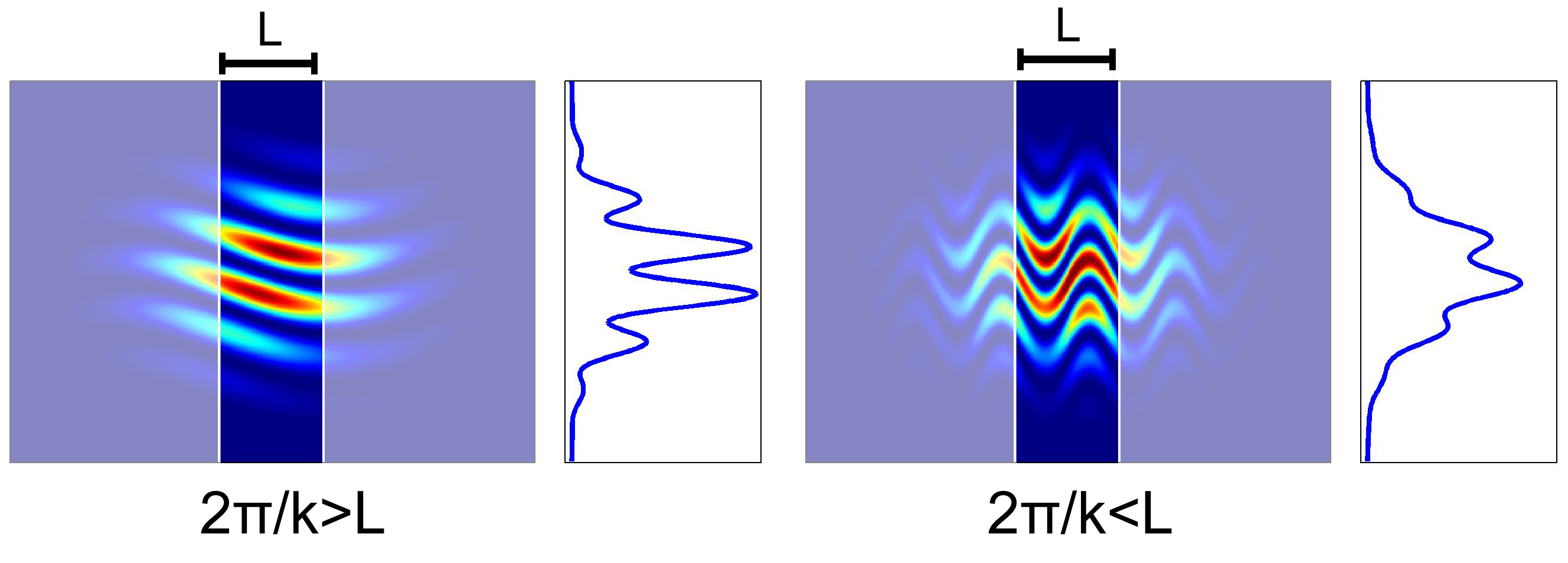}
\caption{Illustration of the filter behavior of the choice of integration length during detection. 
If only one $k$-mode was populated this would lead to the sinusoidal varying pattern displayed in the figures. Large $k$-modes as shown on the left side affect the phase within the integration length in a uniform fashion almost not affecting the contrast. On the other hand high $k$-modes as shown on the right produce a strong variation of the phase within $L$, strongly diminishing the contrast.}
\label{integration length}
\end{figure}

The evolution of $\hat{\phi}(y)$ can be described in Fourier space by a set of decoupled harmonic oscillators of collective modes with momentum $k$. A collective mode with momentum $k$ modulates the relative density and relative phase $\hat{\phi}(y)$ along the condensate in a sinusoidal fashion on a length scale of $\sim {1/k}$. This is visualized in figure~\ref{integration length}.
 
In the experiment the splitting is performed fast in comparison to the timescale for the spread of perturbation which is set by the inverse chemical potential $t_\mathrm{split}<\xi _\mathrm{h}/c=1/\mu$. Thus there is no time for the atoms to develop correlations along the longitudinal trap axis such that the relative density fluctuations $\braketone{\hat{n}^2}_{t_e=0} = \rho/2$  are completely random and not affected by atomic interactions. In particular, there is no correlation between modes with different momenta. Also, for each atom the decision of going to either half of the split system is random and uncorrelated with other atoms, leading to a binomial distribution of atom number fluctuations in each small segment of the 1d system. The respective width of the relative phase distribution follows from the Heisenberg uncertainty relation. Initially, at $t_e=0$, the relative phase fluctuations are governed completely by the atomic shot noise and are practically negligible at length scales $>\xi _\mathrm{h}$. In other words, $\braketone{\hat{\phi}^2}_{t_e=0}\sim 0$ and its fluctuations are strongly suppressed. On the contrary, initial density fluctuations exhibit for $|k| \lesssim \xi _\mathrm{h}^{-1}$ a large excess of noise compared to the zero-temperature state of two split quasi-condensates. In the latter, stationary case the relative density fluctuations are suppressed by atomic repulsion. As a result, almost all the energy is initially distributed in the density fluctuations, with each collective mode $k$ containing equal energy but different populations that scale as 1/$|k|$ as the energy per quanta in each mode is $\propto c |k|$. In particular, this leads to the weaker effects of modes with higher momenta on the evolution of the system. 

During the dynamics of the system the energy of each harmonic oscillator mode $k$ oscillates between the fluctuations in density and fluctuations in phase, driven by interactions between the atoms. This results in a harmonic time dependence of the fluctuation amplitude with a period $\propto 1/{c|k|}$. For short evolution times all phase fluctuation amplitudes grow in magnitude, which leads to a scrambling of the relative phase $ \hat{\phi}(y)$ along the longitudinal direction. This results in an initial rapid decrease in the interference contrast $C(L)$. For longer times, the oscillations in different $k$-modes dephase and the system reaches a quasi-steady state which is thermal in its appearance. In particular, correlation functions take a form that is algebraically equivalent to their respective counterparts in equilibrium. The effective temperature of the quasi-steady state corresponds to the energy which was equally introduced to all $k$-modes by the splitting process. In analogy to the equipartition theorem~\cite{Huang1987} one can deduce that the quasi-steady state can be described by an effective temperature $\Teff$ given by~\cite{Kitagawa2011}
\begin{equation}
k_B \Teff=\frac{\mu}{2}=\frac{\gOneD \rho}{2} \label{Teff_formula},
\end{equation} 
where $\mu$ is the chemical potential, $g$ is the 1d interaction strength and $k_B$ is Boltzmann's constant. As shown in~\cite{Kitagawa2011} the detailed form of the FDFs of $C(L)$ and $C^2(L)$ can then be obtained by sampling the amplitude of the fluctuations of the phase with Gaussian statistics. The choice of a particular integration length $L$ during the detection process represents a filter for the effects of different $k$-modes (figure \ref{integration length}). For an integration length $L$, modes with wavelengths $2\pi/k<L$ produce strong fluctuations of $\hat{\phi}(y)$ within $L$, while modes with wavelengths $2\pi/k>L$ uniformly affect $\hat{\phi}(y)$ in $L$ and just change the overall phase of the interference pattern. This means that the contrast $C$ of the integrated interference pattern is governed by the dynamics of modes with wavelength $2\pi/k<L$.  

The higher the population of a particular $k$-mode, the greater the mean amplitude of its phase fluctuation. For \textit{long} integration lengths $L$, there are many significantly populated modes with $k>2\pi/L$ which affect the integrated interference contrast $C$. With increasing evolution time $t_e$, the dynamics of these large amplitude modes leads to a strong decrease in the probability of observing  a high contrast $C$ and the FDF of the contrast takes the form of an exponential decay.  This is the contrast decay regime. For \textit{short} integration lengths $L$, the contrast $C$ is influenced only by sparsely populated modes with high momenta so the FDF of the contrast preserves a non-zero peak for the whole evolution time.  This is the phase diffusion regime. These two regimes have been experimentally studied in detail in~\cite{Kuhnert2012}. The observation of these two regimes is a direct and intuitive visualization of the multimode nature of 1d Bose gases~\cite{Widera2008, Kuhnert2012}. 

Note that the above discussion is only valid in the case of a perfect symmetric splitting. As discussed in detail in reference~\cite{Kitagawa2011}, a density imbalance between the left and right wells leads to a coupling between the relative and common degrees of freedom. In this case, the initial temperature of the unsplit cloud, which defines the initial state of the common degrees of freedom, influences the dynamics of the relative degrees of freedom and hence the dynamics of the interference pattern. However, the treatment in terms of collective excitations of the system remains valid along with the distinction between phase diffusion and contrast decay dynamics depending on the integration length $L$. The decay of the contrast then happens in two steps. First there is an initial rapid decay analogous to the symmetric splitting case. Secondly, the quasi-steady state which is reached, very slowly evolves in time due to the coupling between relative and common modes\,\cite{Kitagawa2011}. This is, however, not expected to lead to true thermalization as the system remains integrable and the modes will not fully equilibrate. 


\subsection{Quantitative Comparison Between Experiment and Theory}

\begin{figure}
    \centering
        \includegraphics[width=0.8\textwidth]{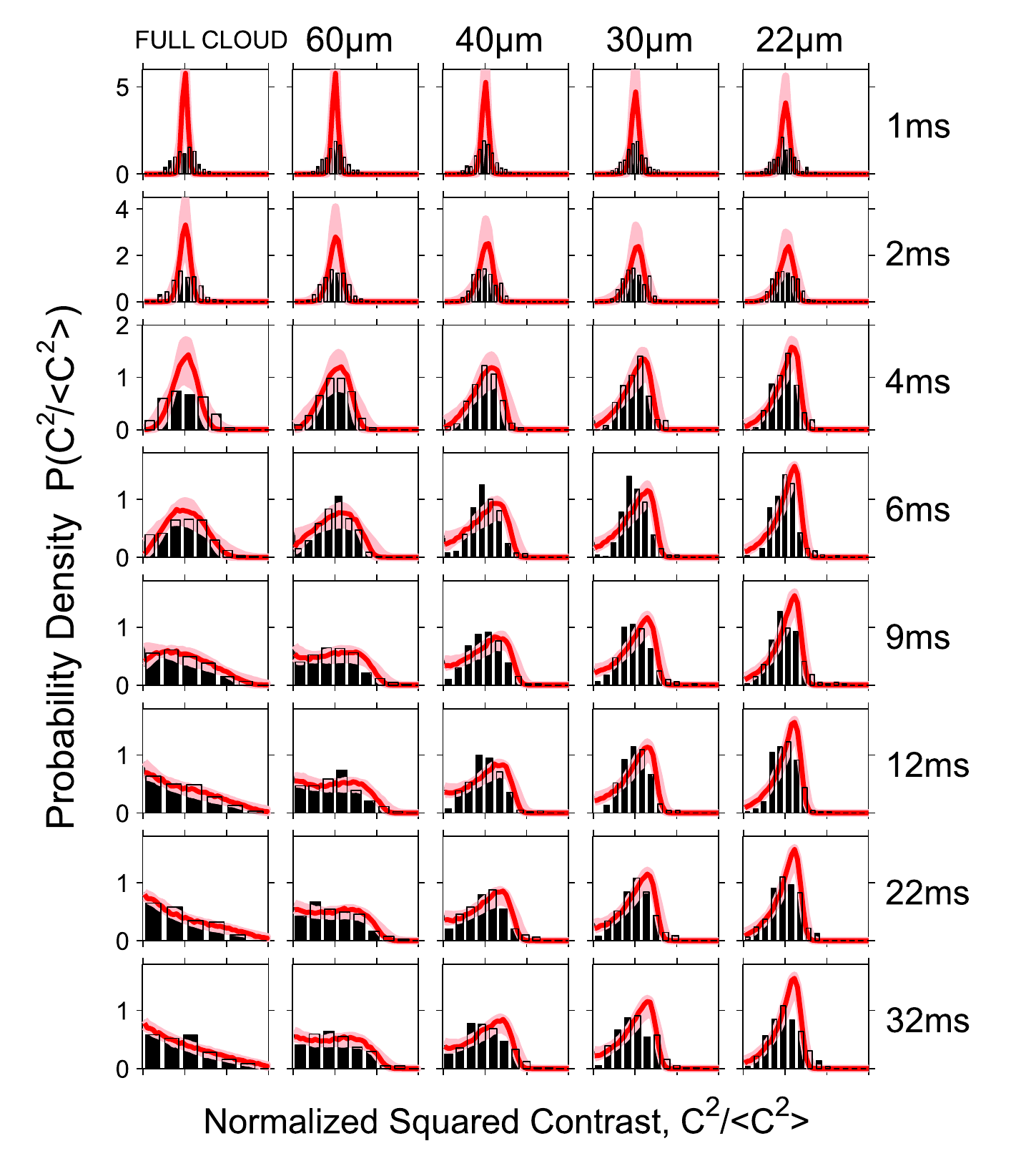}
    \caption{Comparison of distributions of the normalized squared contrast $C^2/\braketone{C^2}$ obtained in section \ref{Pretherm} to the theoretical description of the rapidly split quasi-condensate presented in the previous section \ref{Summary of the theoretic model}. The experimental data is plotted using histograms and the theoretical simulations solid red lines. The light red shaded areas denote the errors resulting from the uncertainty of the experimentally measured theory input parameters. These input parameters were the experimentally measured values of density in a single well $\rho=(32\pm4)\,$atoms/$\mu$m, trap frequency $\omega_\perp=(1.4\pm0.1)\,$kHz, imbalance $\Delta n$ given in percent as $(0.1\pm 0.7)$\%, temperature of the unsplit system of $T=(78\pm 10)\,$nK and the uncertainty of $\pm0.5\,$ms for the point in time of $t_e=0$.
    }
    \label{fig:c2distribution}
\end{figure}

A quantitative comparison between experiment and the Tomonaga-Luttinger liquid model is presented in figure \ref{fig:c2distribution}. We find good agreement of the theoretical description with the experimental data without any free fitting parameter. 

Considering that the theory neglects the effect of the longitudinal trapping potential and contains no fit parameter, only the experimentally measured input parameters, the agreement of experiment and theory is very good. For the first two evolution times the description is, however, less accurate for two reasons. First, the extraction of the contrast is very sensitive for the early evolution times, leading to a broadening of the experimental distributions by fitting uncertainties. Secondly, the disagreements also indicates that the initial preparation of the system through coherently splitting a single cloud is not fast enough to be instantaneous and perfectly uniform as assumed in theory. 

\begin{figure}[ht]
\centering
\includegraphics[width=\textwidth]{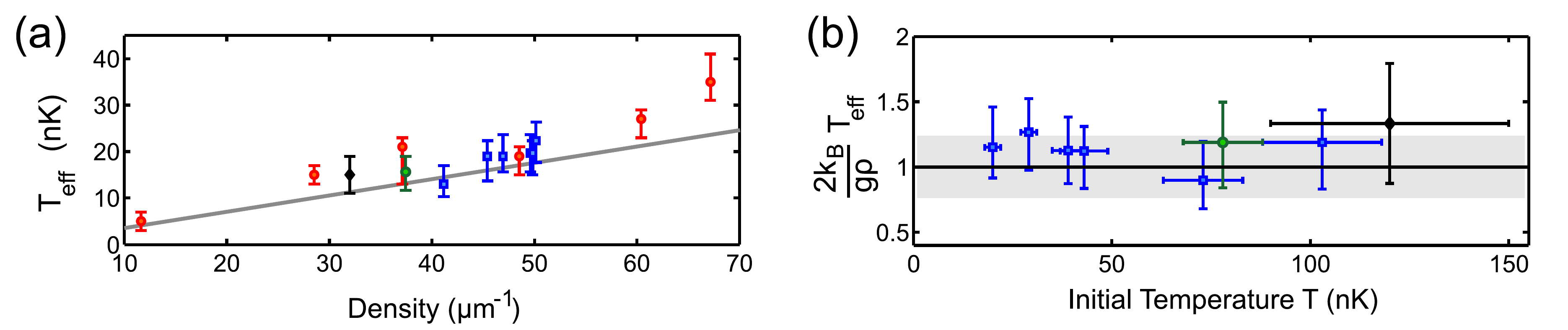}
\caption{a) Dependence of $\Teff$ on density $\rho$ and b) Independence of $\Teff$ from the initial temperature $T$ of the system before splitting, corrected for the scaling of $\Teff$ with density. The (black) solid line corresponds to the theoretical prediction $k_B \Teff= g \rho/2$. The black (green) data point correspond to the dataset presented in section \ref{Pretherm} (\ref{Longterm}) respectively. The figure was taken from~\cite{Gring2012}.}
\label{TeffScaling}
\end{figure}

By preparing 1d Bose gases with a wide range of different parameters, we can further systematically check the predictions of our theoretical model. Figure \ref{TeffScaling} shows the predicted scaling (Equation~\ref{Teff_formula}) of the effective temperature with density, as well as the independence of $\Teff$ from the initial temperature before splitting. 

These two observations provide further strong support for our theoretical model and the interpretation of the observations presented in this work as prethermalization. Especially the observation that the properties of the quasi-steady state are independent from the initial temperature of the unsplit system is another indicator that the observed state is clearly different from the thermal equilibrium state. The observed linear scaling of $\Teff$ with density $\rho$ further shows that the prethermalized state is solely defined by the quantum noise associated with the splitting process. The apparent small systematic offset of the experimentally obtained $\Teff$ and the theoretical description visible in figure \ref{TeffScaling}a can be attributed to imperfections in the experimental splitting process\,\cite{Langen2013}.

Returning to the discussion of the theory in section\,\ref{Summary of the theoretic model},  one sees that  the relative degrees of freedom $ \phi$ and $n$ do indeed decouple from the common degrees of freedom $\phi_\mathrm{com}$ and $n_\mathrm{com}$, at least on the timescales on which we observe the system.  This is particularly interesting, since there is no a priori reason to assume that the different degrees of freedom should not couple and equilibrate in our system. 

\section{Dephasing, not Thermalization} \label{dephasing}

In an earlier experiment by Hofferberth et al. \textit{Nature} 449, 324 (2007)\,\cite{Hofferberth2007a}, the dynamics of a coherently split 1d Bose gas was analyzed through the decay of the coherence factor $\Psi(t)=\braketone{\textit{Re} \int{e^{i \hat{\phi}_s(z,t)}} dz}$. The coherence factor was first introduced in\,\cite{Bistritzer2007} and is closely related to the mean contrast. Note that while the observable $C(L)$ is more easily extracted from experiments, it is more complicated to calculate theoretically. A characteristic sub-exponential decay of the coherence factor  $\Psi(t) \propto \exp[-(t/t_{0})^{\alpha}]$ (with $\alpha \sim 2/3$) was observed, as predicted by Burkov et al.~\cite{Burkov2007} for the thermalization of such a system, where $t_{0}$ is the characteristic timescale for the thermalization. This observation was interpreted as evidence for thermalization. 

The basic physical system investigated in~\cite{Hofferberth2007a} was very similar to that in the new experimental setup presented in this paper. However, at the time of~\cite{Hofferberth2007a}, neither the experimental methods nor the theoretical understanding was sufficiently developed to make a full analysis and characterization of the many-body state to which the system decayed.  As demonstrated in the present work, this characterization was now possible through the new, more complete theoretical description of the non-equilibrium processes~\cite{Kitagawa2010,Kitagawa2011} and the ability to experimentally measure the FDFs which enabled us to characterize the state to which our system decays. We would like to revisit here this old work~\cite{Hofferberth2007a} in the light of our new results. 

In the work of Hofferberth et al.~\cite{Hofferberth2007a} the coherent splitting in was done in the vertical direction, along the direction of gravity. This allowed the direct observation of the phase fluctuations along the length of the system (as with the direct imaging system in this work, but without having to deal with the problem of the chip surface being in the line-of-sight of the imaging beam). However, in general, this simplified imaging scenario comes at the price that a vertical splitting process is not very well controlled and non-symmetric due to the interplay between the pull of gravity and the $1/r$ dependence of the amplitude of the RF radiation emitted from the atom chip wires. As an illustration, a $1\,\mu$m height difference corresponds to change in potential energy of more than 2\,kHz, which is more than the chemical potentials in the 1d gases.  This gravitational sag can be partially compensated by adjusting the polarization of the RF fields, by changing the distance of the trap to the chip surface and by taking advantage of the 1/r dependence of the trapping and RF fields used to generate the double-well potential~\cite{VanEs2008}. However, this constrains the double-well splittings that can be implemented. Furthermore, the compensation is never perfect and variations of the potential depth during the spitting process remain that are on the order of the chemical potential of the 1d gases. 

As a consequence, significant shot-to-shot variations in the atom number difference between the two wells were observed in the previous work~\cite{Hofferberth2007a}.  These overall atom number fluctuations led to an additional phase evolution $\Delta\phi(t)=\frac{t}{\hbar} \Delta \mu$, where $\Delta \mu$ is the difference in chemical potential between the two wells in each experimental run~\cite{Schumm2005a,Lewenstein1996,Javanainen1999}.  Consequently, in~\cite{Hofferberth2007a}, these common fluctuations were rejected by evaluating the coherence factor through $\Psi(t)=\braketone{ \textit{Re} \int{e^{i (\phi(y,t)-\Delta\phi(t))} dy}}=\braketone{\left | \int{e^{i \phi (y,t)} dy} |\right} $. Here  $\Delta\phi(t)$ was determined for each individual interference pattern via $\Delta\phi(t)=\arg{\left(\int dy\, e^{i \phi(y,t)}\right)}$, which for an infinitely long system makes the expression for the coherence factor equivalent to the original formulation of Burkov et al.\,\cite{Burkov2007}. This way of analyzing the coherence factor, however, also rejected the contribution to the dynamics coming from the $k=0$ mode (i.e. the phase evolution which, in a semi-classical picture, can be interpreted as being induced by the quantum shot-noise of the difference between the \textit{total} number of particles in each well).  For a finite size system this $k=0$ contribution to the phase evolution is suppressed only by $1/\sqrt{N}$ where $N$ is the total number of particles in both wells~\cite{Bistritzer2007}.


In addition, in~\cite{Hofferberth2007a}, a strong longitudinal breathing mode was observed after the splitting which limited measurements to the first $11\,$ms of the evolution. As discussed in section~\ref{Experimental details} we split horizontally, perpendicular to the direction of gravity in the new experimental setup presented here. In this way, the symmetry during the splitting process is easily conserved, which makes the initial non-equilibrium state much better defined and the final evolution more repeatable. The atom chip in the new setup is designed in such a way that the longitudinal confinement can be varied independently of the radial confinement. This makes it possible to choose combinations that allow the splitting of a single trap into a double-well with a minimized excitation of longitudinal breathing modes. This made the probing of long evolution times possible.

Furthermore, in the new experimental setup, in comparison to~\cite{Hofferberth2007a}, we split much further away from the atom chip surface, facilitating a double-well separation that is much greater for a given barrier height than in the experimental setup detailed in~\cite{Hofferberth2007a}. Simulations suggest that this leads to a 10 times faster break down of the tunnel coupling between the wells during the splitting process because, due to the larger separation, the overlap of the wave-functions is much smaller once the classical connection between the wells breaks down. Hence, the requirement of fast symmetric splitting to reach the initial state which allows the observation of prethermalization (discussed in~\cite{Kitagawa2010,Kitagawa2011}) is better met in this work. 

Any comparison to current theoretical models of the fast initial evolution requires a clear starting point of the evolution as the exact value of the characteristic exponent $\alpha$ is very sensitive to the position of $t_e=0$. By analyzing the new data and from novel insight from recent experiments on tunnel-coupled systems~\cite{Betz2011}, it became clear that it is very difficult to determine the exact point in time when the two clouds separate during the splitting, the dynamics of the different momentum modes during the final moments of the splitting, and when an independent evolution starts. In the present evaluation we used similar criteria as in~\cite{Hofferberth2007a} to set the starting time of the evolution (simulated plasma frequency $\gg$ 10ms and an observable phase drift). We find that the starting point of the evolution is very difficult to define to better than $\pm$ 0.5\,ms, given our current experimental precision and theoretical understanding.  This is not a problem for observing and analyzing the FDFs and the long-term behavior, but it prevents us from reliably evaluating $\alpha$ from our data. However, for completeness, taking the first 9\,ms of the evolution presented in section \ref{Longterm} and analyzing the coherence factor in the same manner as it was performed in~\cite{Hofferberth2007a}, we extract a value for $\alpha$ that is compatible with the previous result. Yet, studying the system for different initial temperatures, we observe an independence of the characteristic relaxation time $t_0$ from temperature, in stark contrast to the prediction of Burkov et al.~\cite{Burkov2007}. For an ideal, sudden splitting, our model predicts an exponential decay with $\alpha=1$~\cite{Kitagawa2011,Bistritzer2007}. In the experiment the time for the splitting is finite.  Thus, to settle the very interesting question of how the observed dynamics starts and to study its details will require extensive additional theoretical work on the dynamics of the splitting process itself, as well as advanced experimental techniques~\cite{Hohenester2007}.  

In view of the present experiment and the theoretical analysis, the interpretation of Hofferberth et al. \textit{Nature} 449, 324 (2007)~\cite{Hofferberth2007a} should therefore be revised as showing the fast integrable dephasing of the 1d systems, and not thermalization, as was suggested in~\cite{Hofferberth2007a} by the comparison to Burkov et al.\,\cite{Burkov2007}. This revised view is in line with the observation of the absence of thermalization in 1d gases in optical lattices\,\cite{Kinoshita2006}. We point out that for the present experiment, even independent of our theoretical model, the observed clear difference between the FDFs of normalized squared contrast obtained for an equilibrium system formed by cooling in the double-well (figure\,\ref{fig:c2equilibrium}) and the FDFs observed after dynamic splitting (figure\,\ref{fig:c2equilibriumAllfit}) provides direct experimental evidence that no thermalization is observed in our experiment.

\section{Conclusion} \label{Conclusion}
We have detailed the experimental observation of the non-equilibrium many-body phenomenon of prethermalization. This was accomplished by observing the evolution of a rapidly and coherently split one-dimensional Bose gas. Through the use of full quantum mechanical probability distribution functions (FDFs) of the contrast of matter wave interference patterns, the quasi-steady state to which the system decays was found not to be a thermal equilibrium state. Using theory developed for equilibrium systems, we were able to show that the FDFs of the interference contrast in the quasi-steady state are thermal-like in form, but display an effective temperature that is independent of the initial temperature of the gas before the splitting process. The early-time evolution of our system is well described by a Tomonaga-Luttinger liquid model that describes the dephasing of the system to the prethermalized state. This evolution can be visualized by means of FDFs and then contrasted to the true thermalization expected at long times. Moreover, these results clearly illustrate the power of this new method of probing the dynamics of many-body quantum systems through FDFs. Beyond the present work, FDFs are a very general and widely applicable method that can be extended to other quantum many-body systems~\cite{Gerving2012}.

The nature of the true thermalization that manifests itself as a slow evolution towards a steady state after the initial rapid decay to the prethermalized state still remains an open question. A possibility would be the integrability violation through virtual~\cite{Mazets2010,Tan2010,Mazets2011} or even real~\cite{Mazets2010} excitations of the radial degrees of freedom. 

In fact, the question of thermalization in our system is closely connected to the question of how two quantum mechanically correlated but spatially separated objects can lose their memory of that initial correlations, i.e. how do classical properties emerge from a closed quantum system? To answer these questions of thermalization and the emergence of classicality is an ongoing experimental and theoretical endeavor.

\section*{Acknowledgements}
We acknowledge fruitful discussions with Ehud Altman, J\"urgen Berges, Thomas Gasenzer and Anatoli Polkovnikov. Our work was supported by the Austrian Science Fund (FWF) through the Wittgenstein Prize and the EU through the integrating project AQUTE. MK, TL and MG thank the FWF Doctoral Programme CoQuS (\textit{W1210}), RG is supported by the FWF through the Lise Meitner Programme M 1423, I.E.M. acknowledges the financial support from the FWF (project P22590-N16).  DAS acknowledges the EU (grant $N^{o.}\,$\textit{220586}). TK and ED thank DARPA, Harvard-MIT CUA, NSF (Grant No. DMR-07-05472) and ARO-MURI on Atomtronics. 

\section*{References}

\bibliography{NEMBD,phd}
\bibliographystyle{unsrt}


\end{document}